%% file: 0-main.tex
\documentclass[sigconf,nonacm]{acmart}

\input{macro}

\title{Cold-Start Pricing of LLM Services under Two-Sided Learning}
\author{Zhicheng Du}
\email{duzhicheng@ruc.edu.cn}
\affiliation{%
  \institution{Renmin University of China}
  \city{Beijing}
  \country{China}}

\author{Pinyan Lu}
\email{lu.pinyan@mail.shufe.edu.cn}
\affiliation{%
  \institution{Huawei Technologies Ltd}
  \country{China}}
\affiliation{%
  \institution{Shanghai University of Finance and Economics}
  \city{Shanghai}
  \country{China}}

\author{Keyu Wang}
\email{2025200746@ruc.edu.cn}
\affiliation{%
  \institution{Renmin University of China}
  \city{Beijing}
  \country{China}}

\author{Zihe Wang}
\email{wang.zihe@ruc.edu.cn}
\affiliation{%
  \institution{Renmin University of China}
  \city{Beijing}
  \country{China}}

\author{Xiang Yan}
\email{xyansjtu@163.com}
\affiliation{%
  \institution{Huawei Technologies Ltd}
  \country{China}}

\author{Chunxue Yang}
\email{cxueyoung@gmail.com}
\affiliation{%
  \institution{Huawei Technologies Ltd}
  \country{China}}

\renewcommand{\shortauthors}{Du et al.}

\begin{document}

\begin{abstract}

Online large language model (LLM) services face a pricing cold start: providers initially lack the historical data and user information needed for precise pricing, while users' willingness to pay evolves with repeated use.
We formulate this challenge as a \emph{two-sided learning} problem between a provider and a user.
The user learns from private usage experience, whereas the provider incurs a serving cost and dynamically adapts prices using only purchase decisions to maximize cumulative profit.
The provider's central challenge is to learn to price against an evolving willingness to pay without observing the private experiences driving that evolution.

We establish matching upper and lower bounds on the minimax regret over horizon $T$, identifying two cost-dependent regimes.
When serving costs are sufficiently low, our Calibration-and-Tracking (CAT) algorithm achieves the optimal rate $\Theta(T^{1/3})$.
In the higher-cost regime, where the provider must learn whether continued service is profitable, Calibration-and-Tracking with Retirement (CAT-R) attains the optimal rate $\Theta(\sqrt{T})$.
This phase transition shows that serving costs affect not only profit margins but also the fundamental difficulty of learning to price LLM services.
Simulations support the predicted regret scaling and show that our algorithms outperform standard online learning baselines in the evaluated settings.
\end{abstract}

\begin{CCSXML}
<ccs2012>
<concept>
<concept_id>10010405.10010455.10010460</concept_id>
<concept_desc>Applied computing~Economics</concept_desc>
<concept_significance>500</concept_significance>
</concept>
<concept>
<concept_id>10002951.10003260.10003282.10003550</concept_id>
<concept_desc>Information systems~Electronic commerce</concept_desc>
<concept_significance>300</concept_significance>
</concept>
<concept>
<concept_id>10003752.10003809.10010047.10010048</concept_id>
<concept_desc>Theory of computation~Online learning algorithms</concept_desc>
<concept_significance>300</concept_significance>
</concept>
</ccs2012>
\end{CCSXML}

\ccsdesc[500]{Applied computing~Economics}
\ccsdesc[300]{Information systems~Electronic commerce}
\ccsdesc[300]{Theory of computation~Online learning algorithms}

\keywords{LLM service pricing, two-sided learning, minimax regret}

\maketitle

\section{Introduction}
\label{sec:intro}
\input{1-intro}

\section{Preliminaries}
\label{sec:model}
\input{2-model}

\section{Always-Service Market}
\label{sec:optimal-regret}

\input{3-low-cost}

\section{Selective-Service Market}
\label{sec:big-cost}
\input{4-high-cost}

\section{Simulation Experiments}
\label{sec:experiments}

\input{5-experiments}

\section{Conclusion}
\label{sec:conclusion}
\input{6-conclusion}

\bibliographystyle{ACM-Reference-Format}
\bibliography{reference}

\newpage
\appendix
\input{A-proofs}

\end{document}

%% file: macro.tex
\usepackage{fix-cm}
\usepackage{graphicx}
\usepackage[titletoc]{appendix}
\usepackage{xcolor}
\usepackage{color-edits}
\addauthor[Zhicheng]{dzc}{blue}
\addauthor[Codex]{codex}{red}
\makeatletter
\@ifclassloaded{acmart}{}{\usepackage{amsmath,amssymb,amsthm}}
\makeatother
\usepackage{xifthen}
\usepackage{stackengine}
\usepackage{comment}
\usepackage[utf8]{inputenc}
\usepackage{nicefrac}  
\usepackage{xfrac}
\usepackage{enumerate}
\usepackage{CJKutf8} 
\usepackage{algorithm}
\usepackage{algpseudocode}
\usepackage{cleveref} 
\makeatletter
\renewcommand*{\theHALG@line}{\thealgorithm.\arabic{ALG@line}}
\makeatother
\usepackage{subcaption} %
\usepackage{bbm}
\usepackage{natbib}

\usepackage{enumitem}

\theoremstyle{plain}
\newtheorem{assumption}{Assumption}

\newtheorem{theorem}{Theorem}

\newtheorem{definition}{Definition}

\newtheorem{lemma}{Lemma}

\newtheorem{proposition}{Proposition}
\newtheorem{remark}{Remark}

\newcommand{\squishlist}{
\begin{list}{{\small$\bullet$}}{%
\setlength{\itemsep}{0pt}%
\setlength{\parsep}{3pt}%
\setlength{\topsep}{3pt}%
\setlength{\partopsep}{0pt}%
\setlength{\leftmargin}{1.5em}%
\setlength{\labelwidth}{1em}%
\setlength{\labelsep}{0.5em}%
}}
\newcommand{\squishend}{\end{list}}

\newcommand{\expect}[2][]{\mathbb{E}\ifthenelse{\not\equal{}{#1}}{_{#1}}{}\!\left[{\def\givenn{\middle|}#2}\right]}
\newcommand{\indicator}[2][]{\mathbf{1}\ifthenelse{\not\equal{}{#1}}{_{#1}}{}\!\left\{{\def\givenn{\middle|}#2}\right\}}
\newcommand{\Variance}[2][]{\text{Var}\ifthenelse{\not\equal{}{#1}}{_{#1}}{}\!\left[{\def\givenn{\middle|}#2}\right]}

\newcommand{\pubHis}[1]{\mathcal{H}_{#1}^{P}}

\newcommand{\xhdr}[1]{\vspace{8pt}\noindent{\bf {#1.}\ }}

\newcommand{\callOutcome}[1]{X_{#1}}
\newcommand{\callCount}[1]{N_{#1}}
\newcommand{\successCount}[1]{S_{#1}}

%% file: 1-intro.tex
\begin{figure}[t]
    \centering
    \includegraphics[width=\columnwidth,trim=30bp 65bp 30bp 100bp,clip]{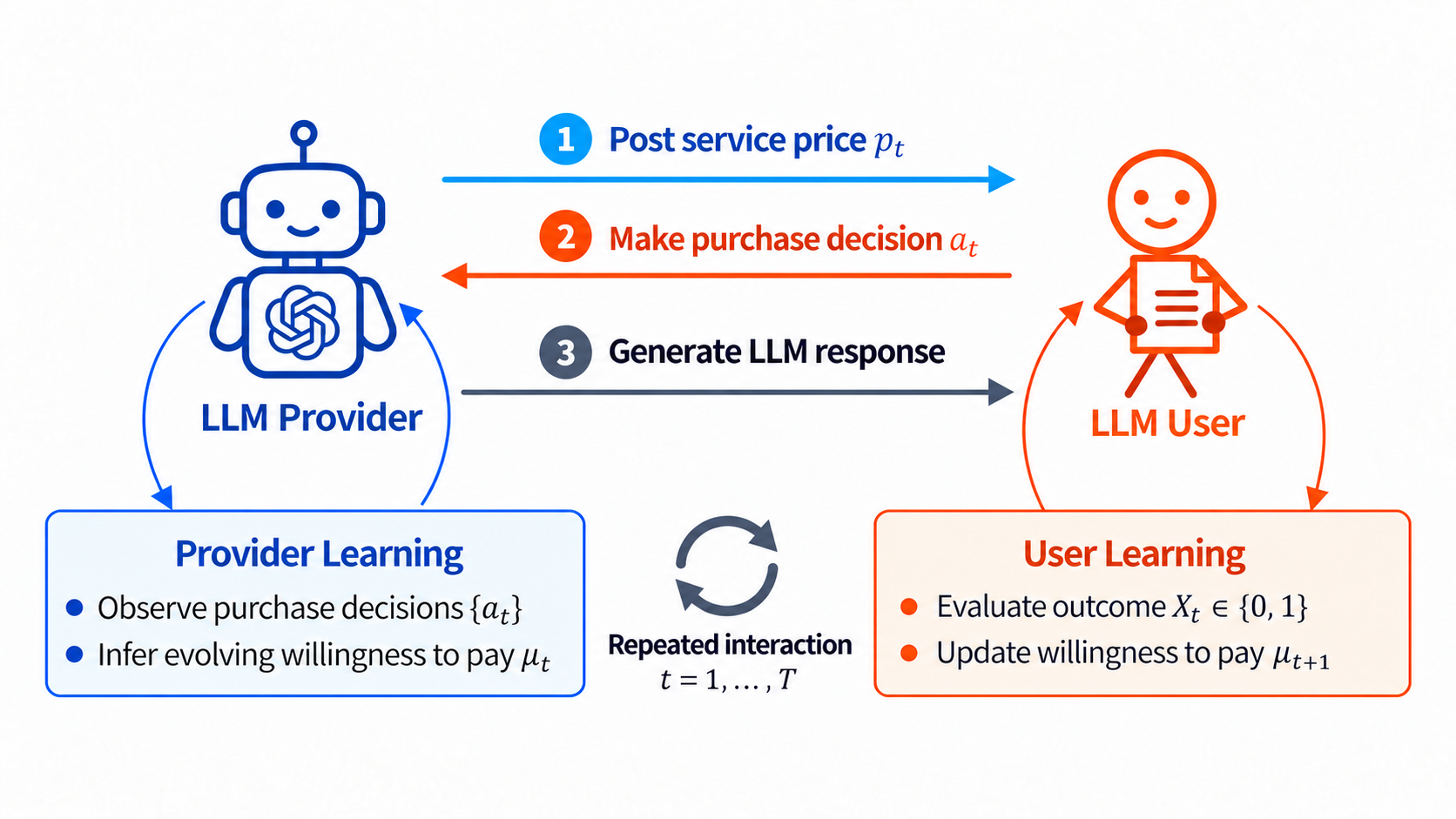}
    \caption{Two-sided learning in online LLM services.}
    \Description{Diagram of repeated interaction between an LLM provider and a user. The provider posts a price, the user makes a purchase decision, and an accepted call generates a response. The provider infers willingness to pay from purchase decisions, while the user privately evaluates task outcomes and updates her willingness to pay.}
    \label{fig:two-sided-learning-model}
\end{figure}

Rapid advances in LLMs have brought LLM-based agents into work and everyday life.
Online LLM services are typically offered through APIs and other pay-per-use interfaces.
They measure and bill usage in tokens, with per-token prices shaping provider profits and user spending.
In 2026, Google reported that its model APIs were processing nearly 22 billion tokens per minute.\footnote{Google, ``Q2 2026 earnings call: Remarks from our CEO,'' \url{https://blog.google/company-news/inside-google/message-ceo/alphabet-earnings-q2-2026/}.}
Token consumption has reached an enormous scale and continues to grow rapidly.

At this scale, even slight price differences can substantially affect user spending and provider profits.
These economic stakes motivate a shift from coarse pricing to more precise, personalized prices in online LLM markets.
As use becomes routine, users have stronger incentives to cautiously weigh the value of each call against its price.
Their willingness to pay can also change with experience.
Providers must therefore track changes in demand and willingness to pay to tailor prices and implement finer price discrimination.

Indeed, providers face a \emph{pricing cold-start} problem.
When launching a new model or service, they often lack relevant transaction histories and user data.
They must learn from repeated market interactions and adjust prices over time.
A key obstacle is that providers observe purchase decisions but not users' private evaluations of task outcomes or the value they obtain.
Yet these evaluations determine users' willingness to pay and hence the provider's profit potential.
Private usage experience also changes willingness to pay over time.
Providers must therefore learn to price against a moving target without directly observing the driving force.

To study and address this pricing challenge, we model the interaction as a \emph{two-sided learning} problem.
A persistent match type, initially unknown to both sides, specifies the probability that a call solves the user's task.
After each call, the user privately observes a binary outcome and updates her belief about the match type.
The provider incurs a serving cost and adjusts prices using only purchase decisions to learn the user's evolving willingness to pay.
Each posted price is therefore both a revenue choice and a binary query about the user's private posterior.
Figure~\ref{fig:two-sided-learning-model} illustrates this repeated interaction and the information available to each side.
From the provider's perspective, we ask how to design a dynamic pricing policy that maximizes long-run profit.

We compare every online policy with a dynamic full-information benchmark that knows the realized match type and observes the private outcomes.
We use minimax regret over horizon $T$ to measure learning difficulty and algorithmic performance.
The serving cost determines two qualitatively different markets.
In both markets, we establish matching upper and lower bounds on minimax regret, revealing a phase transition in its optimal rate.
The optimal regret is $\Theta(T^{1/3})$ in the \emph{always-service market}, where serving remains profitable throughout.
It increases to $\Theta(\sqrt{T})$ in the \emph{selective-service market}, where the provider must also learn whether to retire.
We also conduct Monte Carlo simulations to evaluate finite-horizon regret and demonstrate our algorithms' advantages over standard online pricing baselines in the tested settings.

To summarize, our contributions are threefold.
\begin{enumerate}
    \item \textbf{Modeling pricing cold starts.}
    To our knowledge, we are the first to model and study pricing cold starts in online LLM service markets as a two-sided learning problem.
    The model captures the distinctive economics of LLM services.

    \item \textbf{Regret phase transition.}
    We establish matching upper and lower bounds on minimax regret, obtaining $\Theta(T^{1/3})$ in the always-service market and $\Theta(\sqrt{T})$ in the selective-service market.
    Our CAT and CAT-R algorithms attain these rates by alternating calibration and tracking, with CAT-R adding a retirement test.
    This transition shows that serving costs affect the fundamental difficulty of learning to price.

    \item \textbf{Simulation and implications.}
    Our simulations support the predicted regret scaling and show that our algorithms outperform standard online pricing baselines in the tested settings.
    These findings support alternating calibration and tracking as a design principle for practical dynamic pricing.
\end{enumerate}

\subsection{Related Work}
\label{subsec:related-work}
Our work relates to two strands of literature.

\xhdr{LLM Service Pricing}
LLM services can be viewed as an emerging form of cloud service, typically billed by token usage but valued for task-dependent output quality \citep{velasco2026overcharging}.
Recent work studies competitive pricing across tasks \citep{mahmood2024pricing}, token allocation and fine-tuning menus \citep{bergemann2025economics}, and data-calibrated pricing in LLM routing markets \citep{guo2026pricing}.

Other work examines billing transparency and repeated service delegation.
\citet{velasco2026overcharging} analyze incentives to misreport token usage, while \citet{cao2026secondbest} design mechanisms against providers that substitute cheaper models or inflate token counts.
These studies inform pricing and incentive design, but do not study the joint feedback structure considered here.
In our model, both parties initially lack knowledge of a persistent service match.
The user learns from private task outcomes, while the provider learns to price using only purchase decisions.
To our knowledge, prior work has not studied pricing cold starts under this two-sided learning structure in online LLM markets.

\xhdr{Online Learning/Pricing}
The stochastic bandit literature provides the basic regret language for online decisions \citep{auer2002finite,bubeck2012regret}.
Online learning is now a broad research field focused on designing sequential policies that maximize cumulative rewards under incomplete information.
Related problems in online decision-making include online matching \citep{karp1990optimal,feldman2009online}, online linear programming \citep{agrawal2014dynamic,li2022online}, and, most relevant here, online pricing.
The seminal work of \citet{kleinberg2003value} establishes regret bounds for learning posted prices from purchase decisions.
Subsequent work studies variants of the online pricing problem \citep{amin2013learning,bergemann2006dynamic,besbes2009dynamic}.

Much of this literature treats buyers' values as fixed or models demand directly.
In our model, the user's willingness to pay evolves through Bayesian updates from private call outcomes, with belief updates occurring only after purchases.
The provider must track this evolving belief using purchase decisions alone.
Standard UCB and Thompson-sampling adaptations \citep{auer2002finite,thompson1933likelihood} are natural empirical baselines, but their usual analyses do not directly yield the regret rates established here (see \Cref{sec:experiments}).

Recent work also studies online learning in nonstationary environments \citep{jia2023smooth,leme2021learning,nguyen2025nonstationary}.
The closest work to ours is that of \citet{leme2021learning}, who study purchase-feedback pricing against evolving valuations. 
Their model directly specifies valuation dynamics, whereas in our model, the user's willingness to pay evolves endogenously through Bayesian updates from private call outcomes, with learning occurring only after accepted calls.
In addition, our model is tailored to online LLM services and explicitly accounts for the provider's serving costs. We further show that these costs induce a phase transition in the optimal regret rate (see \Cref{sec:optimal-regret} and \Cref{sec:big-cost}).

%% file: 2-model.tex
In this section, we introduce the two-sided learning setup and explain how the model relates to online LLM services in practice.

\subsection{The Setup}
\label{subsec:basic-setting}

We consider a two-sided learning problem between an LLM service provider (the seller, he) and an LLM user (the buyer, she) over a horizon of $T$ periods, indexed by $t\in[T]:=\{1,\ldots,T\}$.

In each period, the provider posts a price for an LLM service call, and the user arrives with a task and decides whether to make the call to solve her task.
Both parties learn from the information available and adjust their decisions over time.

At the beginning of the interaction, Nature draws a match type $\theta\in\Theta_\varepsilon:=[\varepsilon,1-\varepsilon]$ where $\varepsilon\in(0,1/2)$, according to a prior $F$.\footnote{Restricting types to an interior interval ensures uniformly nondegenerate outcome uncertainty. At $\theta=0$ or $1$, outcomes are deterministic, and pointwise regret can remain bounded independently of the horizon or even be zero.}
The match type $\theta$ is the probability that a call successfully solves the user's task and therefore measures how well the user's tasks match the LLM service.
It depends on both model capabilities and the user's success criteria. 
Neither party knows it ex ante, and both start from the common prior $F$.
As for the prior, we assume:
\begin{assumption}
\label{ass:regular-general-prior}
The prior $F$ has a density $f$ that is uniformly bounded above and bounded away from zero over $\Theta_\varepsilon$.
\end{assumption}

In each period $t\in[T]$, if the user makes a call, the LLM generates a response. The user evaluates this response according to her own private criteria, yielding a binary outcome $\callOutcome{t}\in\{0,1\}$ satisfying
\[
\callOutcome{t}\mid\theta
\overset{\mathrm{i.i.d.}}{\sim}
\operatorname{Bernoulli}(\theta),
\]
where $\callOutcome{t}=1$ indicates that the response successfully solves the user's task.
The value of a successful task to the user is normalized to one. 
The provider incurs a per-call cost $c \geq 0$ associated with model inference and response generation.
Set the price set $\mathcal P:=[0,1]$. 
If the user makes a call at price $p_t$, her
realized utility is $\callOutcome{t}-p_t$, while the provider's realized profit is
$p_t-c$. If the user does not make the call, both players receive zero payoff
in that period.

\begin{remark}
Our per-call formulation can also accommodate linear per-token billing when realized token usage is unknown to both parties before purchase.
Suppose token usage is i.i.d.\ with a commonly known expectation.
Since both parties are risk-neutral, they base their decisions on expected payoffs.
The corresponding expected per-call payment and serving cost are therefore the per-token price and cost multiplied by expected token usage.
\end{remark}

The serving cost determines whether offering service is always profitable for the provider.
We exclude the degenerate high-cost regime $c\geq1-\varepsilon$ and focus on the following two cases:
\begin{itemize}
    \item \emph{Always-Service Market}: $0\leq c\le\varepsilon$. 
    Since every posterior mean is at least $\varepsilon\ge c$, serving at the highest accepted price is always profitable. 
    This regime captures low-cost LLM services for which the provider has no incentive to retire.

    \item \emph{Selective-Service Market}: $\varepsilon<c<1-\varepsilon$.
    The provider's margin may become negative as the user's belief evolves. 
    This regime captures costly LLM services for which the provider must learn whether to continue serving or retire.
\end{itemize}

The central information friction is that the user learns about $\theta$ directly from private call outcomes, while the provider learns about her evolving posterior indirectly from purchase decisions.
We refer to this interaction as a \emph{two-sided learning} process, consisting of the following two components.

\xhdr{User Learning}
The user learns the unknown match type $\theta$ by updating her belief according to Bayes' rule. 
Given the user's purchase history $\{a_{t'}\mid t'<t\}$ and her privately observed outcome history $\{\callOutcome{t'}\mid t'<t,\ a_{t'}=1\}$, define the number of calls and the number of successful calls before period $t$ as
\[
\callCount{t}
:=
\sum\nolimits_{s<t}a_s,
\qquad
\successCount{t}
:=
\sum\nolimits_{s<t,\ a_s=1}\callOutcome{s}.
\]
Her posterior belief is represented by the CDF $G_t$. 
For each $x\in\Theta_\varepsilon$,
\[
G_t(x)
:=
\frac{
\displaystyle\int_{\varepsilon}^{x}
u^{\successCount{t}}
(1-u)^{\callCount{t}-\successCount{t}}
\,\mathrm dF(u)
}{
\displaystyle\int_{\varepsilon}^{1-\varepsilon}
u^{\successCount{t}}
(1-u)^{\callCount{t}-\successCount{t}}
\,\mathrm dF(u)
}.
\]
Let the user's posterior mean before period $t$ be $\mu_t:=\mathbb E_{G_t}[\theta]$.

The user is risk-neutral and myopic, and we assume that she purchases at
equality. Hence, her rational action is
\[
a_t
=
\mathbf 1\{\mu_t\geq p_t\}.
\]

\xhdr{Provider Learning}
Before posting a price, the provider observes only the public history as defined below
$$
\pubHis{t}
:=
\{(p_{t'},a_{t'})\mid t'<t\}.
$$
After posting $p_t$, the provider observes the user's purchase decision $a_t$. 
This decision serves as a binary comparison between the posted price and the user's current posterior mean: 
$$
a_t=1 \quad\Longleftrightarrow\quad \mu_t\geq p_t.
$$

Let $\mathcal A$ denote the class of possibly randomized, nonanticipating pricing algorithms based on public history and the known parameters.
At each round $t$, an algorithm $A\in\mathcal A$ selects a price
\[
p_t
=
A\bigl(\pubHis{t}\bigr)
\in\mathcal P.
\]
The provider faces a learning--profit tradeoff. 
An informative price helps him learn about the user's posterior but may be
rejected. A lower price encourages trade but sacrifices profit and triggers a
private posterior update.
A dynamic pricing algorithm must therefore balance learning and
earning over time.

\xhdr{Timing}
At $t=0$, Nature draws the match type $\theta$ from the prior $F$. 
Both the provider's and the user's initial beliefs are $F$. 
In each period $t\in[T]$, the interaction proceeds as follows.

\begin{enumerate}
    \item Based on the public history $\pubHis{t}$,
    the provider posts a price $p_t\in\mathcal P$ for an LLM service call.

    \item The user observes $p_t$ and decides whether to purchase based on
    her current posterior belief. Her action is denoted by
    $a_t\in\{0,1\}$, where $a_t=1$ represents a call.

    \item The user's decision determines the subsequent interaction:
    \begin{enumerate}[label=\theenumi.\arabic*]
        \item If $a_t=1$, the user privately observes the outcome $\callOutcome{t}$ and updates her belief according to Bayes' rule.

        \item If $a_t=0$, no call occurs and both parties receive zero payoff.
        The user's belief remains unchanged.
    \end{enumerate}
\end{enumerate}

\subsection{Offline Benchmark and Online Problem}
\label{subsec:offline-online}

We evaluate the online provider against a dynamic, full-information benchmark.
The benchmark knows the realized match type $\theta$ before the interaction and observes every realized call outcome, thereby knowing the user's posterior state throughout.

\xhdr{Offline Benchmark}
To define the benchmark, for any posterior state $(n,s)$ with
$0\leq s\leq n$, let $\mu(n,s)$ denote the corresponding posterior mean.
Let $V_h^{\mathrm{OFF}}(n,s;\theta)$ denote the offline expected profit with $h$ periods remaining. Set $V_0^{\mathrm{OFF}}(n,s;\theta)=0$. For
$h\geq1$, the benchmark satisfies the Bellman equation
\[
\begin{aligned}
V_h^{\mathrm{OFF}}(n,s;\theta)
=
\max\Big\{\,
&V_{h-1}^{\mathrm{OFF}}(n,s;\theta),\\
&\mu(n,s)-c
+\theta V_{h-1}^{\mathrm{OFF}}(n+1,s+1;\theta)\\
&\hspace{3.9em}
+(1-\theta)V_{h-1}^{\mathrm{OFF}}(n+1,s;\theta)
\,\Big\}.
\end{aligned}
\]
The first term corresponds to inducing no call in the current period. 
The second corresponds to posting the highest accepted price, $p^*=\mu(n,s)$, and generating a new outcome. 
Thus, at each state, an optimal policy either continues service at this price or retires by posting $p_t=1$ in every remaining period.
The first-best value from the initial state, conditional on $\theta$, is $V_T^{\mathrm{OFF}}(\theta)
:=
V_T^{\mathrm{OFF}}(0,0;\theta)$.\footnote{See Appendix~\ref{apx:offline} for an exact characterization of the optimal benchmark policy using a finite-horizon break-even index.}

\xhdr{Online Problem}
The online provider does not know the realized type $\theta$ and observes only the public history $\pubHis{t}$. 
Given an online algorithm $A$, the expected profit over $T$ periods, conditional on $\theta$, is
\[
\Pi_T(A;\theta)
:=
\mathbb E_{A,\theta}
\left[
\sum_{t=1}^{T}a_t(p_t-c)
\right].
\]
The conditional regret relative to the offline benchmark is
\[
\operatorname{Reg}_T(A;\theta)
:=
V_T^{\mathrm{OFF}}(\theta)
-
\Pi_T(A;\theta).
\]
We use \emph{uniform minimax regret} over the type space $\Theta_\varepsilon$ as our performance measure.\footnote{We seek guarantees across all match types, whereas Bayesian regret averages over $\theta\sim F$ and may mask losses on difficult types.}
An algorithm $A$ achieves \emph{no-regret} if
\[
\sup\nolimits_{\theta\in\Theta_\varepsilon}
\operatorname{Reg}_T(A;\theta)
=
o(T).
\]
The corresponding \emph{optimal regret} is
\[
\mathcal R_T^{*}(\Theta_\varepsilon)
:=
\inf\nolimits_{A\in \mathcal A}
\sup\nolimits_{\theta\in\Theta_\varepsilon}
\operatorname{Reg}_T(A;\theta).
\]

\subsection{Connection to Online LLM Services}
\label{subsec:llm-services}

Our model is designed to capture key features of online LLM service markets, including the following three aspects.

\xhdr{Unknown Match Type}
In practice, neither the user nor the provider knows ex ante how well an LLM will perform on the user's tasks. 
The user may know her tasks but lacks experience with the model, while the provider knows the model's capabilities but not her full task distribution, private context, or evaluation criteria.
We capture this two-sided uncertainty through the latent match type $\theta$, which represents the success probability of the LLM on this user's tasks. 
Treating $\theta$ as persistent over the interaction horizon reflects a fixed LLM and a stable user task profile.

\xhdr{Private Task Outcomes}
A defining feature of real-world LLM services is that task success is user-specific and privately observed, as it depends on the user's intent, task context, and evaluation criteria, which are unavailable to the provider.
The binary outcome $\callOutcome{t}$ captures this information asymmetry: the user observes whether the task is solved, while the provider must infer service quality from subsequent purchase decisions.

\xhdr{Two-Sided Learning}
Repeated interaction naturally induces learning on both sides of an LLM service.
As the user makes more calls, she gains task-level experience and gradually learns how well the LLM performs on her particular tasks, shaping her future willingness to purchase the service.
Meanwhile, the provider observes the user's purchase decisions and uses this behavioral feedback to infer her current valuation of the service and adapt prices accordingly.
Our model captures this asymmetric two-sided learning process in online LLM services.

%% file: 3-low-cost.tex
In this section, we focus on an \emph{always-service market} with $c\le\varepsilon$. 
Since the user's posterior mean is always at least $\varepsilon$, serving at the highest accepted price always yields a nonnegative margin.
The provider therefore need not consider retirement and can focus on adjusting prices to balance learning and profit.

Our main result characterizes the minimax regret.

\begin{theorem}[Minimax regret in the always-service market]
\label{thm:main-minimax-regret}
Under \Cref{ass:regular-general-prior} and assuming $c\in[0,\varepsilon]$, the minimax regret is
\[
\mathcal R_T^*(\Theta_\varepsilon)
=
\Theta(T^{1/3}).
\]
\end{theorem}

We prove the theorem by establishing matching upper and lower bounds in the next two subsections.

\subsection{Upper Bound}
\label{subsec:upper-bound}
We introduce the \emph{Calibration-and-Tracking} (CAT) algorithm, which achieves $O(T^{1/3})$ regret. 
This matches the lower bound, making CAT minimax rate-optimal in the always-service market.

The provider faces a distinctive challenge in this problem. 
Unlike in stationary online learning settings, the user's willingness to purchase evolves over time, like a moving target. 
This evolution is driven by outcomes observed privately by the user. 
Since the provider cannot observe these outcomes directly, he must infer her changing willingness to pay from her responses to posted prices.

Motivated by this obstacle, we design CAT, whose main component alternates between calibration and tracking phases. 
The high-level idea of the
algorithm is as follows.

\xhdr{Calibration Phase}
Calibration identifies the user's hidden success count through price queries.
Given the public call count, this privately observed count determines the user's willingness to pay and hence the provider's profit potential.

A posted price can be viewed as a binary query about this count: the user accepts the price associated with threshold $q$ exactly when her current success count is at least $q$.
CAT exploits this observation through a threshold search.
Starting from a certified lower bound $b$, it queries $b+1$; after an acceptance, it increases $b$ by one and continues.
The first rejection then certifies that the current success count is exactly $b$.
Thus, calibration converts an occasional rejection into an exact public certificate of the user's hidden posterior state.

\xhdr{Tracking Phase}
Tracking keeps prices close to the user's evolving willingness to pay between calibrations.
It allows CAT to earn revenue without recovering the exact success count after every call, balancing underpricing and rejection risk against calibration costs.

Suppose a calibration phase has just ended before period $t$, revealing the exact state $(N_t,S_t)$.
CAT freezes $(n_0,s_0)=(N_t,S_t)$, estimates the match type by the empirical success rate $\widehat{\theta}=s_0/n_0$, and predicts the success count over the next $O(N_t^{2/3})$ calls.
To account for uncertainty from new outcomes, it follows the predicted trajectory with buffered thresholds, posting prices slightly below the predicted willingness to pay to reduce rejection risk at the expense of underpricing.
If a rejection occurs, CAT recovers the current success count exactly and starts a new tracking phase.
If the epoch is completed, CAT proceeds to the next calibration phase.

\vspace{6pt}
We next describe the algorithm in detail. 
Posting the price $\mu(n,q)$ after $n$ calls is accepted if and only if the user's hidden success count is at least $q$. 
As illustrated in \Cref{alg:posterior-mean-query}, we therefore use $\mu(n,q)$ to query whether
the hidden success count is at least $q$.

\begin{algorithm}[htbp]
\caption{Posterior-Mean Query}
\label{alg:posterior-mean-query}
\algrenewcommand{\algorithmiccomment}[1]{\hfill\textcolor{blue}{$\triangleright$ #1}}
\begin{algorithmic}[1]
\Function{Query}{$q,n,t$}
    \If{$t>T$}
        \State \Return $(\mathsf{stop},n,t)$
    \EndIf
    \State Post $\mu(n,q)$ and observe decision $d\in\{0,1\}$
    \State Update the current round $t\gets t+1$
    \If{$d=1$}
        \State Update the service-call count $n\gets n+1$
    \EndIf
    \State \Return $(d,n,t)$
\EndFunction
\end{algorithmic}
\end{algorithm}

\begin{algorithm}[htbp]
\caption{Calibration-and-Tracking (CAT)}
\label{alg:cat}
\algrenewcommand{\algorithmiccomment}[1]{\hfill\textcolor{blue}{$\triangleright$ #1}}
\begin{algorithmic}[1]
\State Initialize $(t,n,s)\gets(1,0,0)$

\Loop
    \State Freeze the posterior state $(n_0,s_0)\gets(n,s)$
    \State Set the tracking-epoch length $E\gets\max\{1,\lceil n_0^{2/3}\rceil\}$
    \State Set the tracking buffer $h\gets\lceil4\sqrt{E}\rceil$
    \State Set the match type estimate
    $\widehat{\theta}\gets s_0/\max\{1,n_0\}$

    \For{$j=0,\ldots,E-1$}
        \Comment{Tracking}
        \State Set the query threshold
        $q_j\gets
        \max\{0,s_0+\lfloor j\widehat{\theta}\rfloor-h\}$
        \State $(d,n,t)\gets\Call{Query}{q_j,n,t}$

        \If{$d=\mathsf{stop}$}
            \State \Return
        \ElsIf{$d=0$}
            \State Update the certified success count $s\gets q_j-1$
            \State \textbf{break}
        \EndIf
    \EndFor

    \If{$d=1$}
        \State Set the starting point for calibration $b\gets q_{E-1}$

        \Repeat
            \Comment{Calibration}
            \State $(d,n,t)\gets\Call{Query}{b+1,n,t}$

            \If{$d=\mathsf{stop}$}
                \State \Return
            \ElsIf{$d=1$}
                \State Update  the calibration threshold $b\gets b+1$
            \EndIf
        \Until{$d=0$}

        \State Update the certified success count $s\gets b$
    \EndIf
\EndLoop

\end{algorithmic}
\end{algorithm}

\begin{lemma}[Upper bound in the always-service market]
\label{thm:upper}
Under \Cref{ass:regular-general-prior} and assuming $c\in[0,\varepsilon]$,
there is a finite constant $C>0$ such that for every integer $T\geq2$,
\[
\mathcal R_T^*(\Theta_\varepsilon)
\leq
\sup\nolimits_{\theta\in\Theta_\varepsilon}
\operatorname{Reg}_T(\mathrm{CAT};\theta)
\leq CT^{1/3}
=O(T^{1/3}).
\]
\end{lemma}

\begin{proof}[Proof sketch]
Regret comes from underpricing on accepted calls and lost profit from rejections.
The key to controlling both is the tracking length $E$.
Tracking produces $O(\sqrt E)$ expected count error, which posterior sensitivity converts into $O(\sqrt E/n)$ underpricing per call.
Calibration restores exact information, while rejection costs from calibration and early restarts contribute $O(1/E)$ per call.
Increasing the tracking length $E$ reduces calibration frequency but increases underpricing.
Balancing these costs gives\footnote{The notation $a\asymp b$ means that $a/b$ is bounded above and below by positive constants independent of $n$.}
$$
\frac{1}{E}\asymp\frac{\sqrt E}{n}
\qquad\Longrightarrow\qquad
E\asymp n^{2/3}.
$$
The resulting expected amortized regret is $O(n^{-2/3})$ per accepted call.
Summing over at most $T$ accepted calls yields
$O(T^{1/3})$.
\end{proof}

Calibration and tracking can be viewed as exploration and exploitation: calibration identifies the user's current willingness to pay, while tracking uses this information to earn revenue.
When the buyer's willingness to pay is fixed, an explore-then-commit policy can first estimate it and then use that estimate to set prices.
In our problem, service itself changes the pricing target: each accepted call gives the user a new private outcome and updates her willingness to pay.
Because these outcomes remain private, information from the last calibration can become outdated during tracking.
Our CAT addresses this feedback by repeatedly recalibrating and then exploiting the updated information while service continues.

The always-service market reflects a common strategy among LLM providers: improving model capabilities while reducing inference and serving costs through architectural improvements, compression, batching, and hardware optimization \citep{kwon2023efficient,patel2024splitwise}.
When costs are sufficiently low, continued service remains profitable. 

Our CAT also offers a template for practical pricing. 
Tracking corresponds to routine price adjustments that follow users' evolving willingness to pay while limiting rejection risk.
Calibration corresponds to occasional diagnostic offers, such as controlled price increases, that refine demand estimates through observed purchase decisions. Alternating these phases allows providers to earn revenue using current information while periodically updating that information as users learn from experience.

\subsection{Lower Bound}
\label{subsec:lower-bound}

For the matching lower bound of the minimax regret, it suffices to identify an instance on which every feasible online algorithm inevitably incurs regret of order $\Omega(T^{1/3})$.

A natural hard instance is the central type $\theta_0=1/2\in\Theta_\varepsilon$.
The corresponding Bernoulli outcome has variance $\theta_0(1-\theta_0)=1/4$, which is maximal among all $\theta\in[0,1]$. 
The hidden success count therefore has particularly large fluctuations, making it difficult for the provider to track the user's belief from purchase decisions.

\begin{lemma}[Lower bound in the always-service market]
\label{thm:lower-bound}
Under \Cref{ass:regular-general-prior} and assuming $c\in[0,\varepsilon]$,
there exist finite constants $\kappa,K>0$ such that for every integer $T\geq2$,
$$
\mathcal R_T^*(\Theta_\varepsilon)
\geq
\inf\nolimits_A \operatorname{Reg}_T(A;1/2)
\geq
\kappa T^{1/3}-K
=\Omega(T^{1/3}).
$$
\end{lemma}

\begin{proof}[Proof sketch]
Fix $\theta=1/2$ and an arbitrary pricing policy $A$.
Couple the online policy and the benchmark using a common sequence of call
outcomes, indexed by accepted calls: each process consumes the next outcome
whenever a call is made.

Let $W_T$ denote the total underpricing loss on accepted calls and $D_T$
the number of rejected periods. 
Matching accepted calls cancels the serving costs and
leaves $D_T$ benchmark calls unmatched.
At $\theta=1/2$, posterior concentration gives
$$
\operatorname{Reg}_T(A;1/2)
\geq
\mathbb E[W_T+d_cD_T]-O(1),
$$
where $d_c=(1/2-c)/2>0$.
It remains to show that these two losses cannot both be small simultaneously.

Policies with sufficiently many rejections satisfy the lower bound.
For the remaining policies, we divide the late horizon into disjoint blocks of length $L\asymp T^{2/3}$.
This captures fresh uncertainty from private outcomes as the user's willingness to pay evolves.
Within each block, conditional on the block-start information, either a rejection occurs with probability at least $1/2$, giving a constant expected loss, or all prices are accepted with probability greater than $1/2$.
In the latter case, binomial anti-concentration and posterior sensitivity give an expected underpricing loss of $\Omega(L^{3/2}/T)$.
Balancing these blockwise lower bounds gives
$$
1\asymp\frac{L^{3/2}}{T}
\qquad\Longrightarrow\qquad
L\asymp T^{2/3}.
$$
Each block contributes $\Omega(1)$ expected loss. Summing over $\Theta(T/L)=\Theta(T^{1/3})$ disjoint blocks
yields the desired lower bound.
\end{proof}

%% file: 4-high-cost.tex
\providecommand{\CATR}{\mathsf{CAT\text{-}R}}

In this section, we focus on the \emph{selective-service market}, where the serving cost satisfies $c\in(\varepsilon,1-\varepsilon)$.
As the name suggests, service becomes selective: the provider must decide whether to continue serving or retire, as well as how to price each call.

The minimax regret exhibits a phase transition across $c=\varepsilon$, increasing from $\Theta(T^{1/3})$ in the always-service market to $\Theta(T^{1/2})$ in the selective-service market.

\begin{theorem}[Minimax regret in the selective-service market]
\label{thm:big-cost-minimax}
Under \Cref{ass:regular-general-prior} and assuming $c\in(\varepsilon,1-\varepsilon)$, the minimax regret is
$$
\mathcal R_T^*(\Theta_\varepsilon)
=\Theta(\sqrt T).
$$
\end{theorem}

We establish matching upper and lower bounds, focusing on the additional challenges relative to \Cref{sec:optimal-regret}.

\xhdr{Upper Bound}
We adapt CAT into \emph{Calibration-and-Tracking with Retirement} (CAT-R).
Its main addition is a \emph{retirement trigger} before each price query.
After $n$ calls, the provider maintains a public upper bound $U$ on the hidden success count.
Each accepted call increases $U$ by one, while an exact calibration resets it to the identified count.
For $n\geq1$, the provider retires if
$$
\frac{U}{n}+r_T(n)<c,
\qquad
r_T(n):=\sqrt{\frac{\log(T/n)}{n}}.
$$
Here $U/n$ bounds the hidden empirical success rate from above, and the bonus $r_T(n)$ allows for sampling uncertainty.

CAT-R also uses a simpler tracking rule.
Starting from an exact state $(n_0,s_0)$, CAT-R prices using the certified lower success count $s_0$, without predicting new successes as CAT does.
This conservative rule incurs greater underpricing but guarantees acceptance during tracking.
The coarser precision is sufficient because learning whether to continue service already entails a square-root minimax loss, as the lower bound below shows.
See \Cref{alg:selective-query,alg:cst-big-cost}.

\begin{algorithm}[htbp]
\caption{Posterior-Mean Query with Retirement}
\label{alg:selective-query}
\algrenewcommand{\algorithmiccomment}[1]{\hfill\textcolor{blue}{$\triangleright$ #1}}
\begin{algorithmic}[1]
\Function{QueryR}{$q,n,U,t$}
    \If{$t>T$}
        \State \Return $(\mathsf{stop},n,U,t)$
    \EndIf
    \If{$n\geq1$ and
    $U/n+r_T(n)<c$}
        \State Post $p_s\gets1$ for every $s=t,\ldots,T$
        \Comment{Retirement}
        \State \Return $(\mathsf{retire},n,U,t)$
    \EndIf
    \State Post $\mu(n,q)$ and observe decision $d\in\{0,1\}$
    \State Update the current round $t\gets t+1$
    \If{$d=1$}
        \State Update the service-call count $n\gets n+1$
        \State Update the success upper bound $U\gets U+1$
    \EndIf
    \State \Return $(d,n,U,t)$
\EndFunction
\end{algorithmic}
\end{algorithm}

\begin{algorithm}[htbp]
\caption{Calibration-and-Tracking with Retirement (CAT-R)}
\label{alg:cst-big-cost}
\algrenewcommand{\algorithmiccomment}[1]{\hfill\textcolor{blue}{$\triangleright$ #1}}
\begin{algorithmic}[1]
\State Initialize $(t,n,L,U)\gets(1,0,0,0)$
\Loop
    \State Freeze the exact certificate $(n_0,s_0)\gets(n,L)$
    \State Set the tracking length
    $E\gets\max\{1,\lceil\sqrt{n_0+1}\rceil\}$
    \For{$j=0,\ldots,E-1$}
        \Comment{Tracking}
        \State $(d,n,U,t)\gets\Call{QueryR}{s_0,n,U,t}$
        \If{$d\in\{\mathsf{stop},\mathsf{retire}\}$}
            \State \Return
        \EndIf
    \EndFor
    \State Set the calibration threshold $b\gets s_0$
    \Repeat
        \Comment{Calibration}
        \State $(d,n,U,t)\gets\Call{QueryR}{b+1,n,U,t}$
        \If{$d\in\{\mathsf{stop},\mathsf{retire}\}$}
            \State \Return
        \ElsIf{$d=1$}
            \State Update the calibration threshold $b\gets b+1$
        \Else
            \State Update the exact certificate $(L,U)\gets(b,b)$
        \EndIf
    \Until{$d=0$}
\EndLoop
\end{algorithmic}
\end{algorithm}

The additional challenge is to balance excessive service when $\theta<c$ against premature retirement when $\theta>c$.
For $\theta<c$, the public count bounds and
$\sum_{n=1}^{T-1}r_T(n)=O(\sqrt T)$ control the loss from continued service.
For $\theta>c$, a maximal concentration bound limits the expected downward estimation error beyond the retirement bonus to $O(T^{-1/2})$.
Charging this error over at most $T$ remaining periods gives an $O(\sqrt T)$ premature-retirement loss.

CAT-R keeps its pricing and calibration losses within this same square-root scale.
Its fixed lower threshold permits count errors up to $E$, giving $O(E/n)$ underpricing per call, while calibration rejections contribute $O(1/E)$ per call.
Balancing these terms gives
$$
\frac{1}{E}\asymp\frac{E}{n}
\qquad\Longrightarrow\qquad
E\asymp\sqrt n.
$$
The resulting $O(n^{-1/2})$ per-call cost sums to $O(\sqrt T)$.
Combining all the losses above yields the $O(\sqrt T)$ upper bound.

\xhdr{Lower Bound}
The additional difficulty is learning whether the market is worth serving.
To construct hard instances, we consider two nearby types $\theta_+=c+\delta$ and $\theta_-=c-\delta$ that are hard to distinguish but favor opposite long-run service decisions: too many calls under $\theta_-$ can produce losses, while too few under $\theta_+$ mean giving up profitable calls.

Let $n$ denote the total number of calls.
The KL divergence between the distributions of the provider's observations under $\theta_-$ and $\theta_+$ is at most $O(\delta^2\mathbb E_{\theta_-}[n])$.
So distinguishing the types requires at least order $1/\delta^2$ calls.
To construct a hard pair, we choose this information requirement to match the horizon:
$$
\frac{1}{\delta^2}\asymp T
\qquad\Longrightarrow\qquad
\delta\asymp T^{-1/2}.
$$
A testing bound then shows that the policy cannot reliably make few calls under $\theta_-$ and serve nearly every period under $\theta_+$.
Combining this bound with the posterior-deviation estimates, and choosing a suitable constant in $\delta$, gives regret of order $\Omega(T\delta)=\Omega(\sqrt T)$ under at least one of the two types.

\xhdr{Phase Transition}
The phase transition reflects an additional learning task.
In the always-service market, the provider need not learn whether continued service is viable.
Once cost lies inside the type support, nearby match types can imply opposite long-run service decisions, and distinguishing them creates a larger unavoidable loss.
Serving cost therefore affects not only profit margins but also the fundamental difficulty of learning to price.
The transition concerns the asymptotic regret rate for fixed costs; it does not imply a discontinuous jump in finite-horizon profit.

Our algorithm may incur negative margins early on to encourage continued use and learn whether service can be profitable in the long run. This resembles the free trials and introductory discounts that providers offer to attract users. Such offers may support future profits, but continued subsidies may not pay off if users' long-run willingness to pay remains below the serving cost.

This risk highlights the importance of assessing long-run profitability. In the selective-service market, providers must learn which users are worth serving. Market assessment and customer segmentation based on expected profitability are therefore particularly relevant. 
The need to learn whether to continue service, rather than just how to price it, is the additional challenge that drives the regret phase transition.

For a fixed service and user population, reducing serving cost raises the margin on every call at a given price.
If this reduction moves the market into the always-service regime, it also removes the viability-learning burden and lowers the optimal worst-case profit shortfall relative to the benchmark from order $\sqrt T$ to order $T^{1/3}$.
This gives an additional economic rationale for improving model capability while reducing inference and serving costs, consistent with efforts toward cost-efficient LLM service design~\citep{chen2023frugalgpt}.

%% file: 5-experiments.tex
The above sections characterize regret scaling as the horizon grows to infinity. We now use simulations to investigate the following questions about finite-horizon performance in both markets.

\begin{enumerate}
    \item[\textbf{Q1.}]
    \textbf{Regret scaling.}
    Do our algorithms, CAT and CAT-R, exhibit the theoretically predicted $T^{1/3}$ and
    $T^{1/2}$ regret scaling in the always-service and selective-service
    markets, respectively?

    \item[\textbf{Q2.}]
    \textbf{Algorithmic comparison.}
    How do CAT and CAT-R compare with other online pricing policies (such as UCB and Thompson Sampling) under
    the same information constraints in their respective markets?
\end{enumerate}

\subsection{Experimental Setup}
\label{subsec:exp-setup}

We conduct Monte Carlo (MC) simulations to evaluate regret scaling and algorithmic performance \citep{rubinstein2008simulation}. 

\xhdr{Parameters}
We set the prior $F=\operatorname{Unif}[\varepsilon,1-\varepsilon]$ with $\varepsilon=1/3$ and fix the match type $\theta=0.5$.
The cost is $c=0.05$ in the always-service market and $c=0.4$ in the selective-service market. 
We consider horizons $T\in\{2^{15},2^{16},\ldots,2^{25}\}$.
For each algorithm and horizon, we conduct 20 runs, each using a different random seed.
These parameter choices serve as representative instances.

\xhdr{Baseline Algorithms}
We compare against four standard online learning and adaptive pricing baselines. 
To apply the three finite-armed bandit algorithms, we uniformly discretize the continuous price space $\mathcal P=[0,1]$ into $K$ grid prices with $K=\Theta((T/\log T)^{1/3})$.\footnote{This order gives a near-optimal grid size in standard one-dimensional Lipschitz bandit models \citep[Section~4.1]{slivkins2019introduction}. We adopt it as a theory-motivated choice.}

\begin{itemize}
    \item \textbf{Price-UCB.}
    Each grid price serves as an arm.
    After trying each grid price once, the policy chooses the price with
    the largest empirical mean profit plus an exploration bonus
    $\sqrt{2\log t/N_i(t)}$, where $N_i(t)$ counts its previous uses.

    \item \textbf{Price-Thompson.}
    The policy maintains independent Beta distributions for purchase probabilities at grid prices, initialized as Beta$(1,1)$.
    Each period, it samples $q_i$ for each price $p_i$ and selects the price maximizing $(p_i-c)q_i$. 
    It then updates the parameters using the observed purchase decision.

    \item \textbf{Epsilon-Greedy.}
    With probability $0.1$, the policy explores a uniformly sampled grid
    price. Otherwise, it selects an untested price if one remains, or
    chooses the price with the highest empirical mean profit.

    \item \textbf{Hill-Climbing.}
    Starting from the initial price $p_1=0.5$, the policy raises the price after an
    acceptance and lowers it after a rejection, using step size
    $0.05/\sqrt{t}$.
\end{itemize}

\xhdr{Offline Benchmark}
All online policies share the same offline benchmark within each market.
In the always-service market, the benchmark posts the highest acceptable price, the user's posterior mean, each period.
In the selective-service market, the retirement decision follows the Bellman recursion, whose $O(T^2)$ state complexity makes exact computation impractical.
We therefore approximate its expected profit using tight upper and lower bounds, with a guaranteed absolute error below $0.011$ in total expected profit at each tested horizon.

\xhdr{Evaluation Metrics}
In each run, cumulative regret is the offline benchmark's total profit minus the algorithm's total profit.
We measure performance by the mean cumulative regret $\bar R(T)$ over 20 runs.
To examine regret scaling, we estimate the empirical regret exponent by fitting $\log_2\bar R(T)$ against $\log_2 T$ using least squares.
If $\bar R(T)\approx CT^k$ for a constant $C>0$, then
\begin{equation*}
\log_2\bar R(T)\approx\log_2 C+k\log_2 T.
\end{equation*}
Thus, the fitted slope estimates the growth exponent $k$.
To assess stability across random seeds, we show pointwise 95\%
Student-$t$ confidence intervals for the mean as shaded regions.

\subsection{Experimental Results}
\label{subsec:exp-result}

\begin{figure}[t]
    \centering
    \includegraphics[width=\columnwidth]{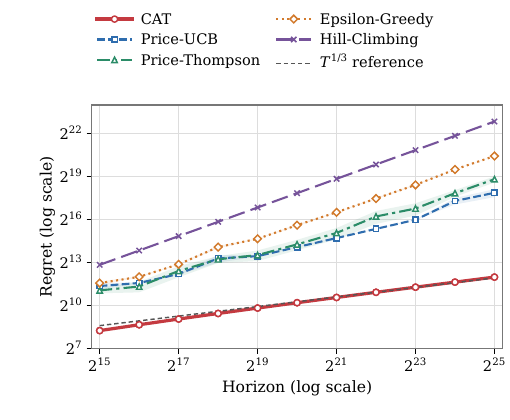}
    \caption{Empirical regret scaling and algorithm comparison in the
    always-service market.}
    \Description{Log-log plot of mean cumulative regret against the time horizon for CAT and four pricing baselines in the always-service market. The red CAT curve is below the baseline curves and is compared with a one-third-power reference line. Shaded regions show confidence intervals.}
    \label{fig:low-cost-regret-scaling}
\end{figure}

\begin{figure}[t]
    \centering
    \includegraphics[width=\columnwidth]{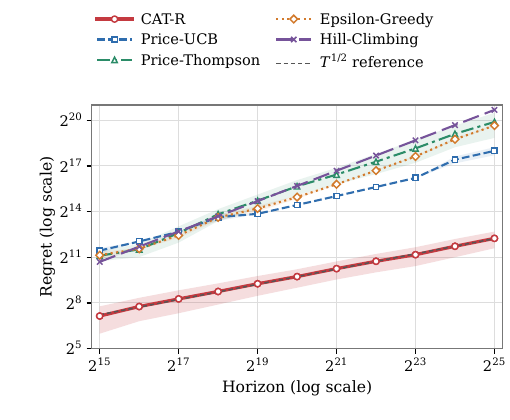}
    \caption{Empirical regret scaling and algorithm comparison in the
    selective-service market.}
    \Description{Log-log plot of mean cumulative regret against the time horizon for CAT-R and four pricing baselines in the selective-service market. The red CAT-R curve is below the baseline curves and is compared with a square-root reference line. Shaded regions show confidence intervals.}
    \label{fig:high-cost-regret-scaling}
\end{figure}

Figures~\ref{fig:low-cost-regret-scaling} and \ref{fig:high-cost-regret-scaling} show regret scaling and algorithmic performance in the always-service and selective-service markets, respectively.
Both figures plot mean cumulative regret against the horizon $T$ on base-2 log--log axes.
Each curve represents an algorithm, with CAT and CAT-R highlighted in red.
Shaded regions show the corresponding confidence intervals, and gray dashed lines mark the theoretical reference rates.
The fitted tail slopes are reported in Table~\ref{tab:regret-tail-slopes}.

\xhdr{Regret scaling}
The red curves exhibit growth close to the predicted $T^{1/3}$ and $T^{1/2}$ rates for CAT and CAT-R, respectively. 
Over the six largest horizons, the red curves are nearly linear on the log--log axes, with least-squares slopes of $0.357$ for CAT and $0.497$ for CAT-R.
The results provide finite-horizon evidence for the different regret scales in the two markets: CAT's fitted exponent remains slightly above $1/3$, while CAT-R's is close to $1/2$.

\begin{table}[tb]
    \centering
    \caption{Fitted regret exponents over $T=2^{20},\ldots,2^{25}$.}
    \label{tab:regret-tail-slopes}
    \begin{tabular}{@{}lcc@{}}
        \toprule
        Algorithm & Always-service & Selective-service \\
        \midrule
        \textbf{CAT} & \textbf{0.357} & --- \\
        \textbf{CAT-R} & --- & \textbf{0.497} \\
        Price-UCB & 0.781 & 0.729 \\
        Price-Thompson & 0.903 & 0.858 \\
        Epsilon-Greedy & 0.973 & 0.951 \\
        Hill-Climbing & 1.000 & 0.999 \\
        \bottomrule
    \end{tabular}
\end{table}

\xhdr{Algorithmic performance}
Our CAT and CAT-R algorithms achieve the lowest mean cumulative regret throughout the experiment. 
At every tested horizon, their red curves lie strictly below all four baselines in both markets.
At $T=2^{25}$, their mean
regrets are approximately $4.07\times10^3$ and $4.84\times10^3$,
compared with $2.40\times10^5$ and $2.60\times10^5$ for Price-UCB.
Table~\ref{tab:regret-tail-slopes} also shows that the baselines have
steeper fitted slopes in both markets.
Price-UCB and Price-Thompson exhibit sublinear empirical growth,
whereas Epsilon-Greedy and Hill-Climbing are close to linear.
Among the baselines, Price-UCB has the
lowest mean regret and Hill-Climbing the highest at every horizon in
this tail range in both markets. 

The advantage of CAT and CAT-R is consistent with their explicit
treatment of two-sided learning. The three bandit baselines follow a
one-sided learning formulation: they pool past observations to estimate
the reward or purchase probability at each price, without explicitly
modeling how the user's private learning changes her willingness to
pay. Hill-Climbing reacts to acceptance and rejection, but likewise
does not reconstruct the user's evolving belief. CAT and CAT-R instead
alternate calibration and tracking, using public purchase decisions
to update information about that belief while earning profit from the
current information. This structure balances information acquisition
and profit as the user learns. The observed performance suggests that
pricing an LLM service can benefit from accounting for users' evolving
valuations as they learn from repeated service calls.

%% file: 6-conclusion.tex
We studied pricing cold starts in online LLM services as a \emph{two-sided learning} problem, where users learn from private experience and providers observe only purchase decisions.
We identify a phase transition in optimal minimax regret driven by serving costs, to our knowledge the first such characterization for LLM service pricing.
In the always-service market, Calibration-and-Tracking (CAT) attains the optimal rate $\Theta(T^{1/3})$.
In the selective-service market, learning whether continued service is profitable raises the optimal rate to $\Theta(\sqrt T)$, attained by Calibration-and-Tracking with Retirement (CAT-R).

These results inform the economic design of generative AI services on the Web, where providers must learn demand while managing serving costs.
Alternating calibration and tracking offers a practical template: occasional diagnostic price offers update demand estimates, while routine price adjustments use those estimates to earn revenue as users learn.
Combining these operations with a retirement test connects personalized pricing with assessments of continued service viability.

Several extensions remain open.
Multi-valued task outcomes would capture different levels of utility and may change both the information needed for pricing and the optimal regret rates.
A fixed positive probability of random purchase decisions would invalidate the exact threshold comparisons used by our algorithms.
The attainable regret rates and conditions for no-regret learning under such noise remain open.
Finally, forward-looking users may strategically accept or reject offers to influence future prices, calling for an analysis that combines learning with incentives.

%% file: A-proofs.tex
\section{Offline Benchmark}
\label{apx:offline}
In \Cref{subsec:offline-online}, we characterize the offline benchmark's optimal policy using a finite-horizon break-even index analogous to the Gittins index \citep{gittins2011multiarmed}.
This characterization holds for any prior supported on $\Theta_\varepsilon$ without requiring the regularity condition in \Cref{ass:regular-general-prior}.

Fix a posterior state $(n,s)$ and a remaining horizon $h\geq1$.
To describe future calls, let $Y_1,\ldots,Y_h$ be independent $\operatorname{Bernoulli}(\theta)$ outcomes, and set the success count
$$
R_i:=s+\sum_{j=1}^{i}Y_j
\qquad \mathrm{with}\qquad R_0=s.
$$
After $i$ further calls, the highest accepted price is $\mu(n+i,R_i)$.

Since rejection provides no new information, an optimal offline policy can serve consecutively at the highest accepted price and then retire. 
The problem therefore reduces to finding an optimal stopping rule.
Let $\mathfrak T_h^0$ denote the stopping times taking values in $\{0,\ldots,h\}$, where $\tau$ specifies the number of calls before retirement. 
Let $\mathfrak T_h^+\subseteq\mathfrak T_h^0$ contain the rules that require at least one call.

\begin{proposition}[Optimal offline index policy]
\label{thm:offline-break-even-policy}
Given a posterior state $(n,s)$, a remaining horizon $h\geq1$, and a match type $\theta$, define the finite-horizon break-even index by
$$
\Gamma_h(n,s;\theta)
:=\max_{\tau\in\mathfrak T_h^+}
\frac{\mathbb E_\theta\!\left[
\sum_{i=0}^{\tau-1}\mu(n+i,R_i)\right]}
{\mathbb E_\theta[\tau]}.
$$
An optimal offline policy posts
$$
p=
\begin{cases}
\mu(n,s), & c\leq\Gamma_h(n,s;\theta),\\
1, & c>\Gamma_h(n,s;\theta).
\end{cases}
$$
In the second case, it retires and posts $p=1$ in all remaining periods.
\end{proposition}

\begin{proof}[Proof of \Cref{thm:offline-break-even-policy}]
We first reduce the pricing decision to two actions.
At state $(n,s)$, all accepted prices induce the same posterior transition, so $p=\mu(n,s)$ maximizes profit conditional on a call. 
All higher prices are rejected and leave the state unchanged; $p=1$ ensures rejection because $\mu(n,s)\leq1-\varepsilon<1$.
Thus only these two actions need be considered. 
The resulting decision tree is finite, so an optimal deterministic policy exists by backward induction; randomization cannot improve its value.

We next express the benchmark value as an optimal stopping problem.
Rejection generates no information and does not change the posterior state.
Hence any deterministic policy using these two actions can be implemented
with the same distribution of call outcomes and total profit by making
its calls consecutively and placing all rejected periods after its last call.
The provider simulates the skipped waiting decisions using
only the outcomes already observed.
The number of calls is then a stopping time in $\mathfrak T_h^0$.
Conversely, every such stopping time is feasible by posting the highest
accepted price until stopping and $p=1$ thereafter. Therefore, the optimal offline value is
$$
V_h^{\mathrm{OFF}}(n,s;\theta)
=\max_{\tau\in\mathfrak T_h^0}
\mathbb E_\theta\!\left[
\sum_{i=0}^{\tau-1}\bigl(\mu(n+i,R_i)-c\bigr)\right],
$$
where the sum is zero for $\tau=0$.

Finally, we use this representation to derive the break-even index rule.
Let $C_h(n,s;\theta)$ be the optimal value among policies that call
immediately. The same argument restricts the stopping time to
$\mathfrak T_h^+$, giving
$$
\begin{aligned}
C_h(n,s;\theta)
&=\max_{\tau\in\mathfrak T_h^+}
\mathbb E_\theta[\tau]
\left(
\frac{\mathbb E_\theta\!\left[
\sum_{i=0}^{\tau-1}\mu(n+i,R_i)\right]}
{\mathbb E_\theta[\tau]}-c
\right),\\
V_h^{\mathrm{OFF}}(n,s;\theta)&=\max\{0,C_h(n,s;\theta)\}.
\end{aligned}
$$
There are finitely many stopping rules on the binary outcome tree, and each
$\tau\in\mathfrak T_h^+$ satisfies $\mathbb E_\theta[\tau]\geq1$.
Thus $C_h$ is positive, zero, or negative depending on whether $c$ is below, equal
to, or above $\Gamma_h$. Calling immediately is optimal when
$c\leq\Gamma_h$, and retirement is optimal when $c\geq\Gamma_h$.
Applying this comparison after each call proves the stated policy by
backward induction, with zero value when no periods remain.
\end{proof}

The index is the largest expected revenue per call before retirement,
analogous to the reward-to-duration ratio of a Gittins index \citep{gittins2011multiarmed}.
In the always-service market,
$\Gamma_h\geq\mu(n,s)\geq\varepsilon\geq c$, so serving in every period is optimal.
In the selective-service market, serving can remain optimal even at a negative current margin when
$\mu(n,s)<c\leq\Gamma_h(n,s;\theta)$, as future profits can offset the current loss.

This distinction determines how the benchmark enters the regret analysis.
In \Cref{apx:pf-regret}, its profit is the sum of all $T$ posterior margins, which can be compared call by call with the online policy.
In \Cref{apx:big cost}, the benchmark may stop at an outcome-dependent call count.
We therefore compare the policies using the same outcome sequence while allowing the benchmark to stop early.

\section{Omitted Proofs in \texorpdfstring{\Cref{sec:optimal-regret}}{Section \ref*{sec:optimal-regret}}}
\label{apx:pf-regret}
\subsection{Proof of the Upper Bound}
\label{apx:pf-upper}

Throughout this subsection, fix an integer $T\geq2$ and a serving
cost $c\in[0,\varepsilon]$.

The proof of \Cref{thm:upper} is organized as follows.
\Cref{lem:posterior-upper-sensitivity} first converts a one-unit difference
in the hidden success count into an $O(1/n)$ difference in the user's
posterior mean. We next record the deterministic large-sample inequalities
used throughout the analysis. \Cref{lem:upper-regret-decomposition} couples the
online process with the benchmark and reduces regret to accepted-call
underpricing plus rejected periods.

For the phase-by-phase analysis, \Cref{lem:upper-invariants} validates the
threshold-query interpretation and every exact reset, while
\Cref{lem:upper-stopped-tail} justifies treating the unused outcomes after an
adaptive reset as a fresh Bernoulli sequence. \Cref{lem:upper-prethreshold}
bounds the losses from the initial one-query epoch and all attempts before the large-sample
threshold.

After this initial range, \Cref{lem:upper-good-start} places every later adaptive attempt in the
concentration region, and \Cref{lem:upper-one-attempt} uses the
preceding ingredients to bound the completion probability and expected
charge of one attempt. \Cref{lem:upper-attempt-count} converts the
per-attempt completion probability into an $O(T^{1/3})$ bound on the number
of attempts. The proof of \Cref{thm:upper} finally aggregates these bounds.

We first record a uniform bound on how much the user's posterior mean can
change when one observed failure is replaced by a success. This bound will be
used throughout the proof to bound the effect of errors in the hidden
success count.

\begin{lemma}[Posterior-mean sensitivity]
\label{lem:posterior-upper-sensitivity}
Under \Cref{ass:regular-general-prior}, there exists a finite constant
$\Lambda_{\varepsilon,F}$ such that, for every call count
$n\geq1$ and success count $0\leq q<n$, the increment in the posterior mean satisfies
$$
0<\mu(n,q+1)-\mu(n,q)
\leq
\frac{\Lambda_{\varepsilon,F}}{n}.
$$
\end{lemma}

\begin{proof}[Proof of \Cref{lem:posterior-upper-sensitivity}]
Fix integers $n\geq1$ and $0\leq q<n$, and let $P_{n,q}$ denote the user's posterior
distribution at state $(n,q)$. Let $V\sim P_{n,q}$. We first compare the
posterior means at states $(n,q)$ and $(n,q+1)$. Define
\[
g(u):=\frac{u}{1-u}.
\]
Since changing the success count from $q$ to $q+1$ multiplies the likelihood
kernel by $g(u)$, the posterior at $(n,q+1)$
is obtained by reweighting $P_{n,q}$ by $g$.
Hence, the increment in the posterior mean satisfies
\begin{align*}
\mu(n,q+1)-\mu(n,q)
&=
\frac{\mathbb E_{P_{n,q}}[Vg(V)]}
{\mathbb E_{P_{n,q}}[g(V)]}
-
\mathbb E_{P_{n,q}}[V]\\
&=
\frac{
\mathbb E_{P_{n,q}}[Vg(V)]
-
\mathbb E_{P_{n,q}}[V]\,
\mathbb E_{P_{n,q}}[g(V)]
}
{\mathbb E_{P_{n,q}}[g(V)]}\\
&=
\frac{
\operatorname{Cov}_{P_{n,q}}(V,g(V))
}
{\mathbb E_{P_{n,q}}[g(V)]}.
\end{align*}
The covariance is strictly positive because $P_{n,q}$ has full support on
$\Theta_\varepsilon$ and $g$ is strictly increasing. 
Moreover, $g$ is
$1/\varepsilon^2$-Lipschitz on $\Theta_\varepsilon$, and
\(g(u)\geq \frac{\varepsilon}{1-\varepsilon}\).
Thus, for independent variables $V,V'\sim P_{n,q}$, the independent-copy
covariance identity\footnote{For independent copies $V,V'$ of a square-integrable
random variable with $h(V)$ also square-integrable,
$\operatorname{Cov}(V,h(V))=\frac12\mathbb E[(V-V')(h(V)-h(V'))]$.}
gives
\[
\begin{aligned}
\operatorname{Cov}_{P_{n,q}}(V,g(V))
&=
\frac12\mathbb E\bigl[(V-V')(g(V)-g(V'))\bigr] \\
&\leq
\frac{1}{2\varepsilon^2}\mathbb E[(V-V')^2]
=
\frac{1}{\varepsilon^2}
\operatorname{Var}_{P_{n,q}}(V).
\end{aligned}
\]

It remains to establish a uniform $O(1/n)$ bound on the posterior variance
$\operatorname{Var}_{P_{n,q}}(V)$. Set $x:=q/n$ and define the
per-observation log-likelihood
\[
\ell_x(u):=x\log u+(1-x)\log(1-u).
\]
For $u\in\Theta_\varepsilon$, we have
\[
-\ell_x''(u)
=
\frac{x}{u^2}+\frac{1-x}{(1-u)^2}
\geq
m_\varepsilon,
\qquad
m_\varepsilon:=(1-\varepsilon)^{-2}.
\]
Let $Q_{n,q}$ be the probability distribution on $\Theta_\varepsilon$ with
density proportional to $\exp\{n\ell_x(u)\}$.
We use $Q_{n,q}$ as a reference distribution whose variance is bounded by the likelihood's curvature.
Its potential
$W(u):=-n\ell_x(u)$ therefore satisfies
$W''(u)\geq nm_\varepsilon$. Applying the one-dimensional Brascamp--Lieb
inequality \citep{brascamp1976extensions}\footnote{On a compact interval, if a density is
proportional to $e^{-W}$ with $W\in C^2$ and $W''>0$, the one-dimensional
Brascamp--Lieb inequality gives
$\operatorname{Var}(h)\leq\mathbb E[(h')^2/W'']$ for absolutely continuous
$h$ with finite right-hand side. No boundary condition on $h$ is required.}
to the identity function $u\mapsto u$ gives
\[
\operatorname{Var}_{Q_{n,q}}(V)
\leq
\mathbb E_{Q_{n,q}}
\left[
\frac{1}{W''(V)}
\right]
=
\mathbb E_{Q_{n,q}}
\left[
\frac{1}{n(-\ell_x''(V))}
\right]
\leq
\frac{(1-\varepsilon)^{2}}{n}.
\]

By \Cref{ass:regular-general-prior}, there are constants
$0<\underline f\leq\overline f<\infty$ such
that $\underline f\leq f(u)\leq\overline f$ for almost every
$u\in\Theta_\varepsilon$. Continuity of $f$ is not needed for the following
comparison.
Since $P_{n,q}$ has density proportional to
$f(u)\exp\{n\ell_x(u)\}$, it is obtained by reweighting
$Q_{n,q}$ by $f$. Therefore, for
every nonnegative measurable function $h$,
\[
\mathbb E_{P_{n,q}}[h(V)]
=
\frac{\mathbb E_{Q_{n,q}}[f(V)h(V)]}
{\mathbb E_{Q_{n,q}}[f(V)]}
\leq
\frac{\overline f}{\underline f}
\mathbb E_{Q_{n,q}}[h(V)].
\]
Let $m_{n,q}:=\mathbb E_{Q_{n,q}}[V]$. 
Using the variational
characterization of variance\footnote{The variational characterization is
$\operatorname{Var}_P(V)=\inf_{a\in\mathbb R}\mathbb E_P[(V-a)^2]$.}
and applying the preceding inequality to
$h(v)=(v-m_{n,q})^2$ yields
\[
\begin{aligned}
\operatorname{Var}_{P_{n,q}}(V)
&\leq
\mathbb E_{P_{n,q}}\bigl[(V-m_{n,q})^2\bigr] \\
&\leq
\frac{\overline f}{\underline f}
\mathbb E_{Q_{n,q}}\bigl[(V-m_{n,q})^2\bigr] \\
&=
\frac{\overline f}{\underline f}
\operatorname{Var}_{Q_{n,q}}(V)
\leq
\frac{\overline f(1-\varepsilon)^{2}}{\underline f}\frac1n.
\end{aligned}
\]

Combining this variance bound with the earlier covariance bound and
using \(
\mathbb E_{P_{n,q}}[g(V)]
\geq
\frac{\varepsilon}{1-\varepsilon}\),
we obtain
\[
0<
\mu(n,q+1)-\mu(n,q)
\leq
\frac{(1-\varepsilon)^3\overline f}
{\varepsilon^3\underline f}\frac1n.
\]
The result follows by setting \(\Lambda_{\varepsilon,F}
:=
\frac{(1-\varepsilon)^3\overline f}
{\varepsilon^3\underline f}\).
\end{proof}

The next ingredients concern the accepted-call outcome sequence and the
phase boundaries generated by CAT. They ensure that every query is feasible,
that a rejection produces an exact certificate, and that the unused outcomes
remain fresh after an adaptive boundary.

For the analysis, define
$$
L_T:=\log(2T),
\quad
B_\varepsilon:=
\max\left\{
512,\frac{16}{\varepsilon^2},
\left(\frac{14}{\varepsilon}\right)^{3/2}
\right\},
\quad
n_\star:=\left\lceil B_\varepsilon L_T^3\right\rceil.
$$
For every accepted count $n\geq0$, also write
$$
E_n:=\max\{1,\lceil n^{2/3}\rceil\},
\qquad
h_n:=\lceil4\sqrt{E_n}\rceil.
$$
Thus the algorithm uses $E=E_{n_0}$ and $h=h_{n_0}$ at a tracking start;
the maximum with one is needed only at the initial state $n_0=0$.
These are abbreviations for the quantities in \Cref{alg:cat}.

\paragraph{Large-sample parameter inequalities.}
The choice of $B_\varepsilon$ ensures that, for every
$L\geq\log4$ and $n\geq B_\varepsilon L^3$,
\begin{equation}
\label{eq:upper-prethreshold-inequalities}
2E_n\sqrt{\frac Ln}\leq\sqrt{E_n},
\qquad
\varepsilon n-2\sqrt{nL}\geq h_n.
\end{equation}
Indeed, $E_n\leq2n^{2/3}$, and the first inequality follows from
$B_\varepsilon\geq512$. For the second inequality, use
$B_\varepsilon\geq16/\varepsilon^2$ and
$B_\varepsilon\geq(14/\varepsilon)^{3/2}$.

The following coupling puts the online process and the benchmark on the same
potential-outcome sequence, indexed by accepted calls. This allows us to
charge a rejected period and any accepted-call underpricing separately.

\paragraph{Accepted-outcome coupling.}
For the analysis, fix an arbitrary $\theta\in\Theta_\varepsilon$ and let
$Y_1,Y_2,\ldots$ be the potential outcomes of successive accepted calls,
where
$$
Y_1,Y_2,\ldots\overset{\mathrm{i.i.d.}}{\sim}
\operatorname{Bernoulli}(\theta).
$$
Write
$$
R_n:=\sum_{i=1}^nY_i,\qquad R_0:=0,\qquad
\mathcal F_n:=\sigma(Y_1,\ldots,Y_n).
$$
If $K_t:=\sum_{r=1}^t a_r$ is the number of calls made by the online
provider through calendar period $t$, then the model's state before period
$t$ is $(N_t,S_t)=(K_{t-1},R_{K_{t-1}})$. A rejected period consumes no
potential outcome.

For analysis, continue CAT indefinitely by omitting only the
$t>T$ stopping check in \textsc{Query}. An \emph{attempt} starts from an exact
certificate and ends at the first rejection: it consists of tracking and,
if all tracking queries are accepted, the subsequent calibration, called
\emph{catch-up} below. The implemented policy consists of the first $T$
periods of this process.

\begin{lemma}[Couple the benchmark and the online process]
\label{lem:upper-regret-decomposition}
For any admissible policy, let $K_T$ be its number of accepted calls by time
$T$, let $\tau_{n+1}$ be the calendar period of its $(n+1)$-st accepted call,
and set
$$
\begin{aligned}
W_T&:=\sum_{n=0}^{K_T-1}\bigl(\mu(n,R_n)-p_{\tau_{n+1}}\bigr),\\
Z_T&:=\sum_{n=K_T}^{T-1}\bigl(\mu(n,R_n)-c\bigr),\qquad
D_T:=T-K_T.
\end{aligned}
$$
Under the always-service condition $0\leq c\leq\varepsilon$,
$$
\operatorname{Reg}_T(A;\theta)=\mathbb E_{A,\theta}[W_T+Z_T]
\leq\mathbb E_{A,\theta}[W_T+D_T].
$$
Moreover, $0\leq W_T+D_T\leq T$ pathwise.
\end{lemma}

\begin{proof}[Proof of \Cref{lem:upper-regret-decomposition}]
The proof proceeds in two steps.

\xhdr{Step 1: Identify the full-information path}
At state $(n,R_n)$ the largest price accepted by the user is
$\mu(n,R_n)$. Since every posterior is supported on
$[\varepsilon,1-\varepsilon]$ and $c\leq\varepsilon$, the corresponding margin
is at least $\varepsilon-c\geq0$. On any fixed potential-outcome path, the
$(i+1)$-st accepted call is made at state $(i,R_i)$ and earns at most
$\mu(i,R_i)-c$. Thus a policy making $m$ calls earns at most
$$
\sum_{i=0}^{m-1}\bigl(\mu(i,R_i)-c\bigr).
$$
This bound depends only on the number of calls, not on their calendar dates,
because a period without a call consumes no outcome. Every summand is
nonnegative, so the bound is maximized at $m=T$ and is attained by calling in
every period at the highest accepted price. Thus an optimal
full-information policy makes $T$ calls and earns
$$
\sum_{n=0}^{T-1}\bigl(\mu(n,R_n)-c\bigr).
$$
Taking expectations identifies this pathwise value with
$V_T^{\mathrm{OFF}}(\theta)$.

\xhdr{Step 2: Align accepted calls}
The online provider's accepted calls consume the prefix
$Y_1,\ldots,Y_{K_T}$. At an accepted call its price cannot exceed
$\mu(n,R_n)$, so $W_T\geq0$. Subtracting the online profit from the
full-information path and adding and subtracting the first $K_T$ benchmark
margins gives the regret identity. The tail $Z_T$ has exactly $D_T$ terms,
and $0\leq\mu(n,R_n)-c\leq1$; hence $0\leq Z_T\leq D_T$. Finally,
$W_T\leq K_T$, so $W_T+D_T\leq K_T+D_T=T$.
\end{proof}

We next verify that the public query answers have the intended threshold
interpretation. This is the invariant that makes every recorded reset exact.

\begin{lemma}[Threshold queries and exact resets]
\label{lem:upper-invariants}
At accepted count $n$, an integer query $q\in\{0,\ldots,n\}$ is accepted if and only if $q\leq R_n$. 
Every calibration, tracking, or catch-up rejection identifies the current success count $R_n$ exactly.
\end{lemma}

\begin{proof}[Proof of \Cref{lem:upper-invariants}]
The proof proceeds in four steps.

\xhdr{Step 1: Verify the threshold interpretation}
By monotonicity from \Cref{lem:posterior-upper-sensitivity},
$q\mapsto\mu(n,q)$ is strictly increasing. Therefore, the tie-breaking rule
and the query price $\mu(n,q)$ give
$$
\mu(n,q)\leq\mu(n,R_n)
\quad\Longleftrightarrow\quad
q\leq R_n.
$$
Because both posterior means lie in $[\varepsilon,1-\varepsilon]$, every query
price is feasible. At $n=0$, $E_0=1$ and the tracking threshold is $q_0=0$,
so the unique initial query is accepted by equality.

\xhdr{Step 2: Establish calibration}
The initial one-query tracking epoch asks $q_0=0$ at $n=0$ and is accepted by
the tie-breaking rule, so the following calibration starts at $n=1$ with
$R_1\geq b=0$. More
generally, at any calibration start the maintained lower bound satisfies
$R_n\geq b$. If $b+1$ is accepted, then the
current count is at least $b+1$, and after the generated outcome the new
count remains at least $b+1$; increasing $b$ by one preserves the invariant.
A rejection of $b+1$ gives $b\leq R_n\leq b$, so $R_n=b$.

It remains to check feasibility. In the initial calibration, after $k$ acceptances the current
values are $b=k$ and $n=1+k$, so the query $b+1=n$ is feasible.
Feasibility of subsequent calibrations is checked in Step 4.

\xhdr{Step 3: Verify tracking resets}
Since $\widehat\theta\in[0,1]$,
$$
q_{j+1}-q_j\in\{0,1\}.
$$
Also $q_0\leq s_0=R_{n_0}$, so the first query cannot be rejected. If $q_j$
is rejected after $q_{j-1}$ was accepted, then
$$
R_{n_0+j}\geq q_{j-1},
\qquad
R_{n_0+j}\leq q_j-1.
$$
The rejection forces $q_j=q_{j-1}+1$ and
$R_{n_0+j}=q_j-1$, which is the recorded certificate. Moreover,
$0\leq q_j\leq s_0+j\leq n_0+j$.

\xhdr{Step 4: Verify catch-up resets}
After tracking completes, the threshold $b=q_{E-1}$ was accepted before the
last tracking outcome, so at the catch-up start $R_m\geq b$, where
$m=n_0+E$. Here $b$ denotes the threshold at the start of
calibration, and $r\geq1$ indexes its queries; the algorithm's updated
threshold is $b+r-1$ before query $r$. The catch-up thresholds increase by one. If the query $b+r$ is
rejected after $r-1$ catch-up acceptances, then
$$
R_{m+r-1}\geq b+r-1,
\qquad
R_{m+r-1}\leq b+r-1,
$$
so the recorded certificate is exact. Feasibility follows from
$b=q_{E-1}\leq n_0+E-1=m-1$, hence
$b+r\leq m+r-1$. All thresholds use only public certificates and public
decisions, so every posted price is public-history measurable.
\end{proof}

The exact-reset invariant also identifies the relevant adaptive phase
boundaries. The next lemma states that the potential outcomes after any such
boundary retain their original conditional law, which is needed for the
one-attempt estimates.

\begin{lemma}[Independence of unused outcomes]
\label{lem:upper-stopped-tail}
Let $\nu$ be the accepted count at any exact reset in the infinite
continuation of \Cref{alg:cat}, and set $\sigma:=\nu+E_\nu$.
Both are almost surely finite stopping times for $(\mathcal F_n)_{n\geq0}$.
For either $\tau=\nu$ or $\tau=\sigma$, the unused outcomes
$Y_{\tau+1},Y_{\tau+2},\ldots$ are i.i.d.\
$\operatorname{Bernoulli}(\theta)$ and independent of the history
through $\tau$.
\end{lemma}

\begin{proof}[Proof of \Cref{lem:upper-stopped-tail}]
The proof includes two steps.

\xhdr{Step 1: Verify the stopping-time property}
Every attempt is almost surely finite. Tracking has finitely
many queries. During calibration, the difference between the true success
count and the maintained lower bound decreases by one after each zero
outcome and remains unchanged after an outcome of one. Since $1-\theta\geq\varepsilon$,
the i.i.d.\ sequence contains infinitely many zeros almost surely, so
calibration ends after finitely many calls.

To identify each boundary, simulate all public thresholds and decisions using
$Y_1,\ldots,Y_n$
until a phase boundary is reached or the next outcome is requested. Rejections
consume no outcome. Thus $\{\nu\leq n\}\in\mathcal F_n$ for each exact reset.
Induction over phase boundaries proves the assertion for all resets.
Since $E_\nu$ is $\mathcal F_\nu$-measurable, decomposing over
$\{\nu=k\}$ also shows that $\sigma=\nu+E_\nu$ is a stopping time.
The tracking decisions are determined by the outcomes available by
$\sigma$, so tracking completion is $\mathcal F_\sigma$-measurable.
On the event that tracking completes, $\sigma$ is the catch-up start count.

\xhdr{Step 2: Prove stopped-tail independence}
Write $\mathcal F_\nu$ for the information available through outcome $\nu$.
For $A\in\mathcal F_\nu$ and a cylinder event $C$ involving the next $r$
outcomes, use $A\cap\{\nu=n\}\in\mathcal F_n$ to obtain
$$
\begin{aligned}
&\mathbb P_\theta\bigl(A\cap C(Y_{\nu+1},\ldots,Y_{\nu+r})\bigr)\\
&\quad=\sum_{n\geq0}\mathbb P_\theta\bigl(A\cap\{\nu=n\}
\cap C(Y_{n+1},\ldots,Y_{n+r})\bigr)\\
&\quad=\mathbb P_\theta(A)\mathbb P_\theta(C).
\end{aligned}
$$
The second equality uses independence of future outcomes from $\mathcal F_n$ at each fixed call count. A monotone-class argument\footnote{The
monotone-class theorem extends an identity verified on cylinder events to the
sigma-field generated by those events.} extends the identity to the
future-tail sigma-field. The same argument applies at $\sigma$,
including on the event that catch-up is reached.
\end{proof}

We next analyze initialization. It includes the initial
one-query tracking epoch, all short tracking phases, and their calibration
queries; its expected charge is linear in the threshold.

\begin{lemma}[Bound pre-threshold initialization]
\label{lem:upper-prethreshold}
Let $C_{\mathrm{pre}}$ be the charge accumulated by CAT before the first
exact reset whose accepted-call count is at least $n_\star$, or before the
horizon if no such reset occurs. Then, uniformly over
$\theta\in\Theta_\varepsilon$,
$$
\mathbb E_\theta[C_{\mathrm{pre}}]
\leq C_{\mathrm{pre}}^{0}(\varepsilon)(n_\star+1)
$$
for a finite constant $C_{\mathrm{pre}}^{0}(\varepsilon)$ independent of
$T$ and $\theta$, where the charge is accepted-call underpricing plus one
unit for each rejected period.
\end{lemma}

\begin{proof}[Proof of \Cref{lem:upper-prethreshold}]
The initial tracking epoch has one query, $q_0=0$, and is accepted, so it
consumes one period and one accepted call. Its charge is at most one. The
following calibration starts at $n=1$ with lower bound $b=0$.

For this calibration, if
$D:=R_1\in\{0,1\}$ is the hidden success count at that point, the number of
accepted calibration queries before the first rejection has conditional mean
$D/(1-\theta)\leq1/\varepsilon$; this is the elementary waiting-time formula
\footnote{If independent Bernoulli trials have failure probability
$1-\theta$, the number of trials required to observe $d$ failures has mean
$d/(1-\theta)$; this is the negative-binomial mean.}
for the number of Bernoulli trials needed to observe $D$ failures. The
final rejection contributes at most one. Truncating at the horizon can only
reduce these nonnegative charges.

It remains to bound the tracking and catch-up phases whose exact reset count
is below $n_\star$. Every such tracking phase starts from an exact state
$(n,s)$ with $n<n_\star$, and its first threshold is at most $s$, so the
first tracking query is accepted. Hence the total number of accepted tracking
queries before the threshold is crossed is at most
$$
n_\star+\max_{1\leq n<n_\star}E_n
\leq 2(n_\star+1).
$$
For a completed phase, let $E$ be its tracking length and let $G$ be the
success-count gap at the start of catch-up. Since
$q_{E-1}\geq s-h_n$ and $R_{n+E}\leq s+E$,
$$
0\leq G\leq E+h_n\leq6E.
$$
The expected number of accepted catch-up queries is therefore at most
$G/(1-\theta)\leq6E/\varepsilon$. Summing over completed pre-threshold
phases gives an $O((n_\star+1)/\varepsilon)$ bound because the sum of their
$E$'s is no larger than the number of accepted tracking queries just bounded.

To include rejection losses, note that there is at most one rejection per calibration or tracking phase, and the
number of such phases is at most one plus the number of accepted tracking
queries. Every accepted-call underpricing loss and every rejection contributes at
most one to the charge. Combining these estimates proves the claim, for
example with $C_{\mathrm{pre}}^{0}(\varepsilon)=20(1+1/\varepsilon)$.
\end{proof}

After the finite pre-threshold range, concentration of the accepted-outcome
sequence gives accuracy guarantees for every adaptive certificate. The next lemma
collects the high-probability event and the deterministic inequalities that
will be used in each later tracking epoch.

\begin{lemma}[Guarantee good adaptive attempt starts]
\label{lem:upper-good-start}
Define
\begin{equation}
\label{eq:upper-global-good}
\mathcal E_T:=
\left\{
|R_m-m\theta|\leq2\sqrt{mL_T}
\text{ for every }1\leq m\leq T
\right\}.
\end{equation}
Uniformly over $\theta\in(0,1)$,
\begin{equation}
\label{eq:upper-bad-probability}
\mathbb P_\theta(\mathcal E_T^c)\leq(2T)^{-7}.
\end{equation}
On $\mathcal E_T$, every main-phase attempt starting at accepted count
$n\geq n_\star$ with exact certificate $s=R_n$ satisfies
\begin{equation}
\label{eq:upper-good-start}
\left|\frac{s}{n}-\theta\right|
\leq\delta_n:=2\sqrt{\frac{L_T}{n}},
\qquad
E_n\delta_n\leq\sqrt{E_n},
\qquad
s\geq h_n.
\end{equation}
In particular, lower clipping in the tracking threshold is inactive.
\end{lemma}

\begin{proof}[Proof of \Cref{lem:upper-good-start}]
For each deterministic $m$, Hoeffding's inequality\footnote{For independent
random variables in $[0,1]$, Hoeffding's inequality gives
$\mathbb P(|R_m-m\theta|\geq x)\leq2\exp(-2x^2/m)$.} gives
$$
\mathbb P_\theta\left(|R_m-m\theta|>2\sqrt{mL_T}\right)
\leq2e^{-8L_T}.
$$
A union bound\footnote{The union bound states that the probability of a finite
union is at most the sum of the probabilities of the individual events.}
proves \eqref{eq:upper-bad-probability}. On $\mathcal E_T$, the first
inequality in \eqref{eq:upper-good-start} holds at every adaptive reset count.
The parameter inequalities in \eqref{eq:upper-prethreshold-inequalities} give
$E_n\delta_n\leq\sqrt{E_n}$ and
$$
s\geq n\theta-2\sqrt{nL_T}
\geq\varepsilon n-2\sqrt{nL_T}
\geq h_n.
$$
This proves all claims.
\end{proof}

The next lemma bounds the tracking completion probability and the total
expected charge of an attempt starting in this good region.

\begin{lemma}[Bounds for one attempt]
\label{lem:upper-one-attempt}
Consider a main-phase attempt starting from an exact state $(n,s)$ with
$n\geq n_\star$, $s=R_n$, and satisfying \eqref{eq:upper-good-start}.
Continue the attempt through its first rejection, even if the horizon
ends earlier.
Let $B_n$ be the event that all $E_n$ tracking queries are accepted,
and let $C_n$ be the total accepted-call underpricing plus one unit
for each rejection.
Uniformly over the start history and $\theta\in\Theta_\varepsilon$,
\begin{align}
\mathbb P_\theta(B_n\mid\mathcal F_n)
&\geq\rho:=\frac{35}{36},
\label{eq:upper-attempt-survival}\\
\mathbb E_\theta[C_n\mid\mathcal F_n]
&\leq C_3,
\label{eq:upper-attempt-charge}
\end{align}
where $C_3>0$ depends only on $(F,\varepsilon)$.
\end{lemma}

\begin{proof}[Proof of \Cref{lem:upper-one-attempt}]
Let $E:=E_n$, $h:=h_n$, and $\widehat\theta:=s/n$.
Write $U_n^{\mathrm{tr}}$ and $U_n^{\mathrm{cu}}$ for the total
accepted-call underpricing during tracking and catch-up, respectively,
with $U_n^{\mathrm{cu}}=0$ when catch-up is not reached.
Conditional on $\mathcal F_n$, define
$$
M_j:=\sum_{i=1}^j(Y_{n+i}-\theta),
\qquad 0\leq j\leq E.
$$
The stopped-tail lemma gives conditionally independent Bernoulli increments.
The good-start conditions make clipping in the tracking thresholds
inactive. Thus, whenever the $j$-th tracking query is reached,
\begin{equation}
\label{eq:upper-tracking-slack}
R_{n+j}-q_j
=M_j+j(\theta-\widehat\theta)
 +(j\widehat\theta-\lfloor j\widehat\theta\rfloor)+h.
\end{equation}

\xhdr{Step 1: Bound the tracking completion probability}
On the event $\max_{j\leq E}|M_j|\leq3\sqrt E$, the right-hand side of
\eqref{eq:upper-tracking-slack} is at least
$-3\sqrt E-E\delta_n+h\geq0$. Thus every tracking query is accepted.
Since
$\mathbb E_\theta[M_E^2\mid\mathcal F_n]
=E\theta(1-\theta)\leq E/4$,
Kolmogorov's maximal inequality\footnote{For a square-integrable
martingale with orthogonal increments, Kolmogorov's maximal inequality gives
$\mathbb P(\max_{j\leq E}|M_j|\geq a)\leq\mathbb E[M_E^2]/a^2$.}
gives
$$
\mathbb P_\theta\left(
\max_{j\leq E}|M_j|>3\sqrt E\mid\mathcal F_n
\right)
\leq\frac{E/4}{9E}=\frac1{36},
$$
which proves \eqref{eq:upper-attempt-survival}.

\xhdr{Step 2: Bound tracking underpricing}
For an accepted tracking query, the sensitivity lemma and
\eqref{eq:upper-tracking-slack} bound its underpricing by
$\Lambda_{\varepsilon,F}(R_{n+j}-q_j)/(n+j)$.
Dropping the indicators that the query is reached and accepted and using
$|\theta-\widehat\theta|\leq\delta_n$ gives
$$
U_n^{\mathrm{tr}}
\leq\frac{\Lambda_{\varepsilon,F}}{n}
\sum_{j=0}^{E-1}\bigl(|M_j|+j\delta_n+h+1\bigr).
$$
By Cauchy--Schwarz\footnote{Cauchy--Schwarz gives
$(\mathbb E|X|)^2\leq\mathbb E[X^2]$ for every square-integrable $X$.},
$\mathbb E_\theta[|M_j|\mid\mathcal F_n]\leq\sqrt j/2$.
Together with $E\delta_n\leq\sqrt E$ and
$h+1\leq6\sqrt E$ for $E\geq4$, this yields
$$
\begin{aligned}
\mathbb E_\theta[U_n^{\mathrm{tr}}\mid\mathcal F_n]
&\leq\frac{\Lambda_{\varepsilon,F}}{n}
\left(
\frac{E^{3/2}}{3}+\frac{E^{3/2}}{2}+6E^{3/2}
\right)\\
&\leq\frac{7\Lambda_{\varepsilon,F}E^{3/2}}{n}
\leq14\sqrt2\,\Lambda_{\varepsilon,F},
\end{aligned}
$$
where $E\leq2n^{2/3}$.

\xhdr{Step 3: Bound the terminal gap}
To bound catch-up underpricing, set $m:=n+E$ and $b:=q_{E-1}$.
Define $G:=R_m-b$ on $B_n$ and $G:=0$ otherwise.
On $B_n$, the final tracking query is accepted, so $G\geq0$.
The definition of $q_{E-1}$ gives
$$
G=M_E+E(\theta-\widehat\theta)+\widehat\theta
 +\bigl((E-1)\widehat\theta
 -\lfloor(E-1)\widehat\theta\rfloor\bigr)+h.
$$
For $E\geq4$, the good-start inequalities and
$h\leq4\sqrt E+1$ imply $G\leq|M_E|+8\sqrt E$ on $B_n$.
Therefore,
\begin{equation}
\label{eq:upper-terminal-gap}
\mathbb E_\theta[G^2\mathbf 1_{B_n}\mid\mathcal F_n]
\leq2\mathbb E_\theta[M_E^2\mid\mathcal F_n]+128E
\leq129E.
\end{equation}
This estimate does not condition on $B_n$.

\xhdr{Step 4: Bound catch-up and total charges}
On $B_n$, let $K_G$ be the number of accepted catch-up queries and
$H_G$ their total success-count gap.
Set both variables to zero outside $B_n$.
After $k$ accepted catch-up queries, define
$$
D_k:=R_{m+k}-(b+k),\qquad D_0=G.
$$
Then $H_G=\sum_{k=0}^{K_G-1}(D_k-1)$.
If $D_k=d\geq1$, the next query is accepted and
$D_{k+1}=d-1+Y_{m+k+1}$.
If $D_k=0$, the next query is rejected, ending the attempt.
In particular, $G=0$ gives $K_G=0$.

The stopped-tail lemma applies at the catch-up start because
$B_n$ is $\mathcal F_m$-measurable.
For $G\geq1$, $K_G$ is the number of trials needed to observe $G$ zeros.
Hence $\mathbb E_\theta[K_G\mid\mathcal F_m]=G/(1-\theta)<\infty$.
At residual level $d$, the success-count gap is $d-1$, and the expected
number of accepted queries at that level is $1/(1-\theta)$.
Summing over $d=1,\ldots,G$ gives
\begin{equation}
\label{eq:upper-catchup-slack}
\mathbb E_\theta[H_G\mid\mathcal F_m]
=\frac{G(G-1)}{2(1-\theta)}
\leq\frac{G^2}{2\varepsilon}.
\end{equation}

Posterior sensitivity converts \eqref{eq:upper-catchup-slack}
into an underpricing bound.
The tower property and \eqref{eq:upper-terminal-gap} then give
$$
\mathbb E_\theta[U_n^{\mathrm{cu}}\mid\mathcal F_n]
\leq
\frac{\Lambda_{\varepsilon,F}}{2\varepsilon n}
\mathbb E_\theta[G^2\mathbf 1_{B_n}\mid\mathcal F_n]
\leq\frac{129\Lambda_{\varepsilon,F}E}{2\varepsilon n}
\leq\frac{129\Lambda_{\varepsilon,F}}{\varepsilon}.
$$
Catch-up ends almost surely because $1-\theta>0$.
Whether rejection occurs during tracking or after catch-up, the attempt
contains exactly one rejection. Thus
$C_n=U_n^{\mathrm{tr}}+U_n^{\mathrm{cu}}+1$ almost surely.
Combining the two underpricing bounds proves
\eqref{eq:upper-attempt-charge} with
$C_3:=14\sqrt2\,\Lambda_{\varepsilon,F}
+129\Lambda_{\varepsilon,F}/\varepsilon+1$.
\end{proof}

We next bound the number of epochs to obtain a total-charge bound. Completed epochs increase the accepted-call count by their full tracking
length, while each attempt with a good start has a uniformly
positive conditional probability of completing tracking.
The next lemma combines these two facts.

\begin{lemma}[Number of good-start attempts]
\label{lem:upper-attempt-count}
For each post-threshold attempt $r$ in the infinite continuation,
let $\nu_r\geq n_\star$ be its starting call count.
Let $I_r$ indicate that the attempt starts by calendar time $T$ and
satisfies
$|R_{\nu_r}-\nu_r\theta|\leq2\sqrt{\nu_rL_T}$.
The number $A_T^{\mathrm g}:=\sum_r I_r$ of these good-start attempts
satisfies
\begin{equation}
\label{eq:upper-good-attempt-count}
\mathbb E_\theta[A_T^{\mathrm g}]
\leq C_A(T^{1/3}+1)
\end{equation}
for a numerical constant $C_A>0$, uniformly over
$\theta\in\Theta_\varepsilon$.
\end{lemma}

\begin{proof}[Proof of \Cref{lem:upper-attempt-count}]
The proof contains two steps.

\xhdr{Step 1: Count completed tracking phases}
Let $M_T$ count tracking phases that start at an accepted count
at least $n_\star$ and have all their queries accepted by calendar
time $T$.
If $m_k$ is the starting call count of the $k$-th such phase, then
$m_{k+1}\geq m_k+E_{m_k}\geq m_k+m_k^{2/3}$.

For $m\geq1$, put $u=m^{-1/3}\in(0,1]$.
Concavity gives
$(1+u)^{1/3}\geq1+(2^{1/3}-1)u$, and hence
$$
(m+m^{2/3})^{1/3}-m^{1/3}\geq2^{1/3}-1.
$$
If $M_T\geq1$, telescoping over consecutive completed phases and
using $m_{M_T}\leq T$ gives
\begin{equation}
\label{eq:upper-successful-count}
M_T\leq1+\frac{T^{1/3}}{2^{1/3}-1}.
\end{equation}
This bound also holds when $M_T=0$.

\xhdr{Step 2: Use conditional survival}
The indicator $I_r$ is $\mathcal F_{\nu_r}$-measurable
because the calendar start time and the exact certificate are
determined by the outcomes available at that start. Let $Z_r$ indicate
that all tracking queries of attempt $r$ are accepted under a natural
continuation, even if the horizon ends before the epoch does.

For these good starts, the one-attempt lemma gives, on $I_r=1$,
$\mathbb E_\theta[Z_r\mid\mathcal F_{\nu_r}]\geq\rho$: every attempt starts
with $\nu_r\geq n_\star$, and the remaining two good-start conditions follow
from $I_r=1$ and \eqref{eq:upper-prethreshold-inequalities}. At most one such
natural continuation can be truncated by the horizon, so
$$
\rho\,\mathbb E_\theta\left[\sum_r I_r\right]
\leq\mathbb E_\theta\left[\sum_r I_rZ_r\right]
\leq\mathbb E_\theta[M_T+1].
$$
Together with \eqref{eq:upper-successful-count}, this proves
\eqref{eq:upper-good-attempt-count}, for example with
$C_A:=\bigl(2+(2^{1/3}-1)^{-1}\bigr)/\rho$.
\end{proof}

We now combine the charges from initialization, good post-threshold
epochs, and the exceptional event to prove the uniform upper bound.

\begin{proof}[Proof of \Cref{thm:upper}]
Fix an integer $T\geq2$, a cost $c\in[0,\varepsilon]$, and
$\theta\in\Theta_\varepsilon$.
By \Cref{lem:upper-invariants}, CAT is admissible.
Under the accepted-outcome coupling, let
$\Gamma_T:=W_T+D_T$.
By \Cref{lem:upper-regret-decomposition},
$\operatorname{Reg}_T(\mathrm{CAT};\theta)
\leq\mathbb E_\theta[\Gamma_T]$.
We bound this expected charge in four steps.

\xhdr{Step 1: Bound the initialization charge}
We first separate initialization from post-threshold attempts.
Let $C_{\mathrm{pre}}$ be the charge accumulated through the first
exact reset with accepted count at least $n_\star$, truncated at
time $T$, as in \Cref{lem:upper-prethreshold}.
Write $C_{\mathrm{main}}:=\Gamma_T-C_{\mathrm{pre}}$ for the
remaining charge.
If no post-threshold attempt starts by time $T$, then
$C_{\mathrm{main}}=0$.

The initialization bound gives
$$
\mathbb E_\theta[C_{\mathrm{pre}}]
\leq C_{\mathrm{pre}}^{0}(\varepsilon)(n_\star+1).
$$
This bound includes any tracking or calibration phase that crosses
the threshold before producing an exact reset.
Thus every attempt contributing to $C_{\mathrm{main}}$ starts
from an exact state with accepted count at least $n_\star$.

\xhdr{Step 2: Aggregate the charges of good-start attempts}
We next combine the bounded expected charge of one attempt with
the bound on the number of good-start attempts.
Use $\nu_r$, $I_r$, and $A_T^{\mathrm g}$ from
\Cref{lem:upper-attempt-count}.
For each post-threshold attempt $r$, let $\widetilde C_r$ be its
total charge when continued through its first rejection, even if
this occurs after time $T$.

The attempts starting by time $T$ partition the main-phase periods,
and only the last attempt can be unfinished at the horizon.
Because every charge is nonnegative, continuing an unfinished
attempt can only increase its charge.
Moreover, each such attempt starts at an accepted count
$\nu_r\leq T$.
Hence, on the event $\mathcal E_T$ in
\eqref{eq:upper-global-good}, each such start satisfies
$I_r=1$, and therefore
$$
C_{\mathrm{main}}\mathbf 1_{\mathcal E_T}
\leq\sum_r I_r\widetilde C_r.
$$

To bound the right-hand side, condition on the outcomes available
at each attempt's start.
The indicator $I_r$ is determined by this information.
On $\{I_r=1\}$, its concentration condition and
\eqref{eq:upper-prethreshold-inequalities} imply all the
good-start conditions in \eqref{eq:upper-good-start}.
By \Cref{lem:upper-stopped-tail}, the unused outcomes retain their
i.i.d.\ Bernoulli$(\theta)$ law at this adaptive start.
Thus \Cref{lem:upper-one-attempt} applies and gives
$\mathbb E_\theta[\widetilde C_r\mid\mathcal F_{\nu_r}]
\leq C_3$ on $\{I_r=1\}$.
The tower property\footnote{The tower property allows us to first
take the conditional expectation given the attempt-start history
and then average over that history.} yields
$$
\begin{aligned}
\mathbb E_\theta[I_r\widetilde C_r]
&=\mathbb E_\theta\!\left[
I_r\mathbb E_\theta[
\widetilde C_r\mid\mathcal F_{\nu_r}]
\right]\leq C_3\mathbb E_\theta[I_r].
\end{aligned}
$$
This calculation conditions only on the start history;
it does not condition on $\mathcal E_T$, which also restricts
future outcomes.

All summands are nonnegative, so Tonelli's theorem\footnote{
Tonelli's theorem allows expectations and countable sums of
nonnegative random variables to be interchanged.}
and \eqref{eq:upper-good-attempt-count} give
\begin{equation}
\label{eq:upper-main-charge}
\begin{aligned}
\mathbb E_\theta[
C_{\mathrm{main}}\mathbf 1_{\mathcal E_T}]
&\leq\sum_r\mathbb E_\theta[I_r\widetilde C_r]\\
&\leq C_3\mathbb E_\theta[A_T^{\mathrm g}]\\
&\leq C_3C_A(T^{1/3}+1).
\end{aligned}
\end{equation}

\xhdr{Step 3: Bound the charge on the exceptional event}
When $\mathcal E_T$ fails, we bound the total charge directly,
without estimating individual attempts.
By \Cref{lem:upper-regret-decomposition},
$0\leq\Gamma_T\leq T$ on every outcome sequence.
Combining this with \eqref{eq:upper-bad-probability} gives
\begin{equation}
\label{eq:upper-bad-charge}
\mathbb E_\theta[
\Gamma_T\mathbf 1_{\mathcal E_T^c}]
\leq T\mathbb P_\theta(\mathcal E_T^c)
\leq T(2T)^{-7}.
\end{equation}
Thus the exceptional event contributes only once to the total
bound, regardless of how many attempts occur.

\xhdr{Step 4: Combine the bounds uniformly}
We now combine initialization, the main-phase charge on
$\mathcal E_T$, and the total charge on its complement.
Since $\Gamma_T=C_{\mathrm{pre}}+C_{\mathrm{main}}$ and both
charges are nonnegative,
$$
\begin{aligned}
\mathbb E_\theta[\Gamma_T]
&\leq \mathbb E_\theta[C_{\mathrm{pre}}]
+\mathbb E_\theta[
C_{\mathrm{main}}\mathbf 1_{\mathcal E_T}]
+\mathbb E_\theta[
\Gamma_T\mathbf 1_{\mathcal E_T^c}]\\
&\leq
C_{\mathrm{pre}}^{0}(\varepsilon)(n_\star+1)
+C_3C_A(T^{1/3}+1)
+T(2T)^{-7},
\end{aligned}
$$
where the last inequality uses
\Cref{lem:upper-prethreshold},
\eqref{eq:upper-main-charge}, and
\eqref{eq:upper-bad-charge}.

Finally, $n_\star\leq B_\varepsilon L_T^3+1$ with
$L_T=\log(2T)$, and
$(\log(2T))^3=O(T^{1/3})$ uniformly for $T\geq2$.
The initialization and exceptional-event terms can therefore
be absorbed into a finite constant depending only on
$(F,\varepsilon)$.
Consequently,
$$
\operatorname{Reg}_T(\mathrm{CAT};\theta)
\leq\mathbb E_\theta[\Gamma_T]
\leq C_{F,\varepsilon}T^{1/3}.
$$
All estimates are uniform over $\theta\in\Theta_\varepsilon$
and $c\in[0,\varepsilon]$.
Since CAT is admissible, taking the supremum over $\theta$ and
then using the definition of minimax regret yields
$$
\mathcal R_T^*(\Theta_\varepsilon)
\leq
\sup_{\theta\in\Theta_\varepsilon}
\operatorname{Reg}_T(\mathrm{CAT};\theta)
\leq C_{F,\varepsilon}T^{1/3}.
$$
This proves \Cref{thm:upper}.
\end{proof}

\subsection{Proof of the Lower Bound}
\label{apx:pf-lower}

We first establish the local sensitivity of the posterior mean. 
We then decompose regret into accepted-call underpricing and rejected periods, and translate posted prices into success-count thresholds.
A concentration bound for the success counts and a blockwise tradeoff lemma complete the preparation.

To keep calendar time and accepted-call time distinct, the appendix uses
$K_t$ for the number of accepted calls through period $t$ and $R_n$ for the
number of successes among the first $n$ accepted calls. In the model's notation, before period $t$ we have
$$
\callCount{t}=K_{t-1},
\qquad
\successCount{t}=R_{K_{t-1}}.
$$

Throughout this subsection, fix $\theta=1/2$ and let $Y_1,Y_2,\ldots$ be independent Bernoulli$(1/2)$ potential outcomes.
Set the success count $R_n:=\sum_{i=1}^nY_i$ for every $n\geq0$, with $R_0=0$.

The first lemma quantifies how the posterior mean changes with the
success count. Recall that $\mu(n,s)$ denotes the posterior mean after $n$
calls and $s$ successes.

\begin{lemma}[Posterior sensitivity]
\label{lem:posterior-sensitivity}
Under \Cref{ass:regular-general-prior}, there exist constants $\eta>0$,
$\lambda_F>0$, and an integer $n_F\geq1$ such that
$$
\mu(n,s+1)-\mu(n,s)
\geq
\frac{\lambda_F}{n}
$$
whenever $n\geq n_F$, $0\leq s<n$, and
$|s/n-1/2|\leq2\eta$.
\end{lemma}

\begin{proof}[Proof of \Cref{lem:posterior-sensitivity}]
The proof proceeds in three steps.

\xhdr{Step 1: Localize the central posterior}
Choose $\eta>0$ such that
$[1/2-3\eta,1/2+3\eta]\subset(\varepsilon,1-\varepsilon)$.
Let $P_{n,s}$ denote the posterior distribution at state $(n,s)$ and set
$x=s/n$. Its density is proportional to
$\exp\{n\ell_x(u)\}f(u)$, where
$$
\ell_x(u):=x\log u+(1-x)\log(1-u).
$$ 
For $x$ in the stated central interval,
$\ell_x$ is maximized at $u=x$. Moreover, there is a finite constant $M$
such that, for every $u\in\Theta_\varepsilon$,
\begin{equation}
\label{eq:lower-likelihood-quadratic}
2(u-x)^2
\leq
\ell_x(x)-\ell_x(u)
\leq
M(u-x)^2.
\end{equation}
The first inequality is Pinsker's inequality for Bernoulli distributions,
which lower-bounds the relative entropy between two Bernoulli laws by twice
the squared difference of their means.\footnote{Pinsker's inequality states
that $D_{\mathrm{KL}}(P\|Q)\geq2\,\mathrm{TV}(P,Q)^2$; for Bernoulli laws,
the total-variation distance is the absolute difference of their means.}
The second follows from Taylor's theorem because the interval is bounded away
from zero and one.\footnote{Taylor's theorem with a bounded second derivative
gives a quadratic upper bound on the difference between a smooth function and
its value at a maximizer.}

\xhdr{Step 2: Bound the posterior variance}
By \Cref{ass:regular-general-prior}, we have $0<\underline f\leq f(u)\leq\overline f<\infty$ for almost every $u\in\Theta_\varepsilon$.
By \eqref{eq:lower-likelihood-quadratic}, the normalizing
constant of $P_{n,s}$ is at most
$$
\overline f\exp\{n\ell_x(x)\}
\int_{\mathbb R}\exp\{-2n(u-x)^2\}\,\mathrm du
=
\overline f\sqrt{\frac{\pi}{2n}}\exp\{n\ell_x(x)\}.
$$
For all sufficiently large $n$, the intervals
$$
I_-:=\left[x-\frac{2}{\sqrt n},x-\frac{1}{\sqrt n}\right]
\quad\text{and}\quad
I_+:=\left[x+\frac{1}{\sqrt n},x+\frac{2}{\sqrt n}\right]
$$
are contained in $\Theta_\varepsilon$. Each interval has unnormalized
posterior mass at least
$\underline f\exp\{n\ell_x(x)-4M\}/\sqrt n$. Hence there is a constant
$\rho_F>0$, independent of $(n,s)$ in the central region, such that
$$
P_{n,s}(I_-)\geq\rho_F 
\qquad \mathrm{and} \qquad
P_{n,s}(I_+)\geq\rho_F.
$$

If $V,V'$ are independent draws from $P_{n,s}$, then
$$
\operatorname{Var}_{P_{n,s}}(V)
=
\frac{1}{2}\mathbb E[(V-V')^2]
\geq
\frac{4\rho_F^2}{n}.
$$

\xhdr{Step 3: Bound the posterior-mean sensitivity}
To convert this variance bound into a change in the posterior mean, define
$g(u):=u/(1-u)$. The posterior at $(n,s+1)$ is obtained by
multiplying the density of $P_{n,s}$ by $g$ and renormalizing, and therefore
$$
\mu(n,s+1)-\mu(n,s)
=
\frac{\operatorname{Cov}_{P_{n,s}}(V,g(V))}
{\mathbb E_{P_{n,s}}[g(V)]}.
$$
Since $g'(u)\geq1/(1-\varepsilon)^2$ and
$g(u)\leq(1-\varepsilon)/\varepsilon$ on $\Theta_\varepsilon$, the
independent-copy identity for covariance\footnote{For independent copies
$V,V'$ of a square-integrable random variable, the independent-copy identity
is $\operatorname{Cov}(V,h(V))=\frac12\mathbb E[(V-V')(h(V)-h(V'))]$.}
gives
\[
\begin{aligned}
\operatorname{Cov}_{P_{n,s}}(V,g(V))
&=\frac{1}{2}\mathbb E[(V-V')(g(V)-g(V'))]\\
&\geq\frac{\operatorname{Var}_{P_{n,s}}(V)}{(1-\varepsilon)^2}.
\end{aligned}
\]
Combining these estimates proves the lemma with $\lambda_F:=4\rho_F^2\varepsilon/(1-\varepsilon)^3$.
\end{proof}

The next lemma decomposes regret into the two losses that any policy must limit.
Let $U$ denote the policy's random seed, independent of $Y$.
Unless otherwise stated, probabilities and expectations below average over both $Y$ and $U$.

\begin{lemma}[Regret decomposition]
\label{lem:lower-regret-decomposition}
Under \Cref{ass:regular-general-prior}, fix $c\in[0,\varepsilon]$,
$\theta=1/2$, and an admissible policy $A$. Let $\tau_{n+1}$ be the calendar period
of the $(n+1)$-st accepted call and define
$$
\begin{gathered}
W_T:=\sum_{n=0}^{K_T-1}\bigl(\mu(n,R_n)-p_{\tau_{n+1}}\bigr),
\\
Z_T:=\sum_{n=K_T}^{T-1}\bigl(\mu(n,R_n)-c\bigr),
\qquad
D_T:=T-K_T.
\end{gathered}
$$
Set $d_c:=(1/2-c)/2>0$. There is a finite constant $B\geq0$, depending only on
$(F,c,\varepsilon)$, such that
\begin{equation}
\label{eq:lower-regret-decomposition}
\begin{aligned}
\operatorname{Reg}_T\left(A;\frac12\right)
=\mathbb E_{Y,U}[W_T+Z_T]\geq
\mathbb E_{Y,U}[W_T+d_cD_T]-B.
\end{aligned}
\end{equation}
\end{lemma}

\begin{proof}[Proof of \Cref{lem:lower-regret-decomposition}]
The proof proceeds in three steps.

\xhdr{Step 1: Identify the benchmark path}
At state $(n,R_n)$, the largest price accepted by the user is
$\mu(n,R_n)$. Every posterior mean lies in
$[\varepsilon,1-\varepsilon]$, so the corresponding margin is at least
$\varepsilon-c\geq0$. On a fixed potential-outcome path, a benchmark making $m$
calls earns at most the sum of the first $m$ purchase-indexed margins
$\mu(n,R_n)-c$. Adding one more call leaves the earlier margins unchanged and
adds a nonnegative margin. Thus, an optimal full-information benchmark makes
$T$ calls and earns $\sum_{n=0}^{T-1}\bigl(\mu(n,R_n)-c\bigr)$.

\xhdr{Step 2: Align accepted calls}
The online policy consumes the prefix $Y_1,\ldots,Y_{K_T}$. At each accepted
call, its price is at most $\mu(n,R_n)$, so $W_T\geq0$. On a fixed realization,
define the difference between benchmark and online profits
$$
\Delta_T^{\mathrm{path}}
:=
\sum_{n=0}^{T-1}\bigl(\mu(n,R_n)-c\bigr)
-
\sum_{n=0}^{K_T-1}\bigl(p_{\tau_{n+1}}-c\bigr).
$$
Separating the first $K_T$ benchmark terms from the remaining terms gives
$\Delta_T^{\mathrm{path}}=W_T+Z_T$. Taking expectations yields the equality
in the statement.
This is the same call-by-call coupling as in \Cref{lem:upper-regret-decomposition}.

\xhdr{Step 3: Bound the loss from unmatched calls}
To cover $c=\varepsilon$, we use concentration at $\theta=1/2$
instead of the possibly zero margin $\varepsilon-c$. For
$x=s/n$ with $|x-1/2|\leq\eta$, the likelihood bounds in the proof of
\Cref{lem:posterior-sensitivity}, namely
\eqref{eq:lower-likelihood-quadratic}, imply, for all sufficiently large $n$,
$$
|\mu(n,s)-x|
\leq
\frac{\overline f\int_{\mathbb R}|v|e^{-2nv^2}\,\mathrm dv}
{\underline f\int_{-1/\sqrt n}^{1/\sqrt n}e^{-nMv^2}\,\mathrm dv}
\leq\frac{C_0}{\sqrt n}
$$
for a finite constant $C_0$ depending only on $(F,\varepsilon)$.
Here $v=u-x$, $M$ is the likelihood-bound constant from that proof, and
the denominator integrates over $u\in\Theta_\varepsilon$.
The choice of $\eta$ gives $\eta<(1/2-\varepsilon)/3<d_c$.
Choose an integer $n_0$ so that the bound holds and
$C_0/\sqrt n\leq d_c-\eta$ for every $n\geq n_0$. Then
$|R_n/n-1/2|\leq\eta$ implies $\mu(n,R_n)-c\geq d_c$.

Write $(z)_+:=\max\{z,0\}$. Since $\mu(n,R_n)-c\geq0$,
Hoeffding's inequality gives the finite bound
$$
B:=\sum_{n=0}^{\infty}
\mathbb E_Y\bigl[(d_c-\mu(n,R_n)+c)_+\bigr]
\leq d_cn_0+2d_c\sum_{n=n_0}^{\infty}e^{-2\eta^2n}<\infty.
$$
The tail $Z_T$ contains $D_T$ terms, so pathwise
$$
Z_T\geq d_cD_T-
\sum_{n=0}^{T-1}(d_c-\mu(n,R_n)+c)_+.
$$
Taking expectations proves the claimed inequality. The constant
$B$ does not depend on the policy or the horizon.
\end{proof}

\Cref{lem:lower-regret-decomposition} reduces the lower bound to a comparison between accepted-call underpricing and rejected periods. 
The next lemma expresses an arbitrary posted price as a success-count threshold and quantifies the underpricing caused by a threshold gap.

\begin{lemma}[Convert prices into success-count gaps]
\label{lem:lower-threshold-slack}
Under \Cref{ass:regular-general-prior}, use the constants from \Cref{lem:posterior-sensitivity}.
For $n\geq0$ and $p\in[0,1]$, define
$$
q(n,p):=\min\left(
\{q\in\{0,\ldots,n\}:\mu(n,q)\geq p\}\cup\{n+1\}
\right).
$$
Then a price $p$ is accepted at state $(n,R_n)$ if and only if
$R_n\geq q(n,p)$. Moreover, if $n\geq n_F$,
$|R_n/n-1/2|\leq\eta$, $w$ is a nonnegative integer with
$w\leq\eta n$, and the accepted price satisfies
$R_n-q(n,p)\geq w$, then
$$
\mu(n,R_n)-p\geq\frac{\lambda_Fw}{n}.
$$
\end{lemma}

\begin{proof}[Proof of \Cref{lem:lower-threshold-slack}]
For fixed $n$, increasing the success count from $q$ to $q+1$ multiplies the
posterior density by the strictly increasing likelihood ratio
$u/(1-u)$. The covariance identity used in \Cref{lem:posterior-upper-sensitivity} therefore implies
$\mu(n,q+1)>\mu(n,q)$. The user's tie-breaking rule then gives
$$
p\leq\mu(n,R_n)
\quad\Longleftrightarrow\quad
R_n\geq q(n,p).
$$
If the query is accepted and $R_n-q(n,p)\geq w$, then
$p\leq\mu(n,q(n,p))\leq\mu(n,R_n-w)$. Summing the lower bound in
\Cref{lem:posterior-sensitivity} over the $w$ intervening states gives
$$
\mu(n,R_n)-p
\geq
\mu(n,R_n)-\mu(n,R_n-w)
\geq
\frac{\lambda_Fw}{n}.
$$
Indeed, every intervening count
$R_n-w\leq r<R_n$ satisfies $|r/n-1/2|\leq2\eta$, as required by
\Cref{lem:posterior-sensitivity}.
\end{proof}

The price-loss bound in
\Cref{lem:lower-threshold-slack} is useful only when the hidden success-count path remains
in the central region. We therefore record the high-probability event on which
this is guaranteed at all sufficiently late call counts.

\begin{lemma}[Probability bound for success-count deviations]
\label{lem:lower-central-path}
Let $\eta$ be chosen as in \Cref{lem:posterior-sensitivity}.
Define $\mathcal E_T$ as the event that $|R_m-m/2|\leq\eta m$ for every integer $m\in[\lceil T/3\rceil,T]$.
Then
\begin{equation}
\label{eq:lower-central-probability}
\mathbb P_Y(\mathcal E_T^c)
\leq 2T\exp\left(-\frac{2\eta^2T}{3}\right).
\end{equation}
\end{lemma}

\begin{proof}[Proof of \Cref{lem:lower-central-path}]
For a fixed $m\geq T/3$, Hoeffding's inequality\footnote{For
independent random variables in $[0,1]$, Hoeffding's inequality gives
$\mathbb P(|R_m-m/2|\geq x)\leq2\exp(-2x^2/m)$.} gives
$$
\mathbb P_Y\bigl(|R_m-m/2|>\eta m\bigr)
\leq
2\exp(-2\eta^2m)
\leq
2\exp\left(-\frac{2\eta^2T}{3}\right).
$$
Applying the union bound\footnote{The union bound states that the probability
of a finite union of events is at most the sum of their probabilities.} over
at most $T$ possible values of $m$ proves the claim.
\end{proof}

It remains to quantify the rejection--underpricing tradeoff on one late block using \Cref{lem:lower-threshold-slack,lem:lower-central-path}.
This is the key probabilistic step: either the block contains a
rejection with substantial probability, or private outcome fluctuations force
accepted prices to be too low.

\begin{lemma}[Blockwise loss bound]
\label{lem:lower-block-tradeoff}
Maintain the assumptions and $d_c$ of
\Cref{lem:lower-regret-decomposition}.
Let $I=\{s,\ldots,s+L-1\}$ be a block in the last quarter of the
horizon, with $L=\lfloor T^{2/3}\rfloor$.
Let $W_I$ be the total underpricing on accepted calls in $I$,
and let $D_I$ be the number of rejections in $I$.
There is a constant $b=b(F,c,\varepsilon)>0$ such that,
for all sufficiently large $T$,
\begin{equation}
\label{eq:lower-block-loss}
\begin{aligned}
\mathbb E[W_I+d_cD_I]
\geq b\,\mathbb P(K_{s-1}\geq T/3)-\lambda_F\mathbb P(\mathcal E_T^c).
\end{aligned}
\end{equation}
\end{lemma}

\begin{proof}[Proof of \Cref{lem:lower-block-tradeoff}]
Take $T$ large enough that $L\geq512$.
Write $\mathbb P_s$ and $\mathbb E_s$ for probability and expectation
conditional on the entire interaction before period $s$, including
realized private call outcomes, and on the policy's independent random
seed $U$. This conditioning is used only for analysis.

Under this conditioning, $n:=K_{s-1}$ and $R_n$ are fixed.
For each possible $n$, the event $K_{s-1}=n$ and the history on that
event depend only on $U,Y_1,\ldots,Y_n$.
Thus the unused outcomes remain i.i.d.\ Bernoulli$(1/2)$.
Let $C_I$ be the event that every price in the block is accepted,
and set $Q_I:=\mathbb P_s(C_I)$.

\xhdr{Step 1: Bound loss when a rejection is likely}
If $Q_I\leq1/2$, at least one rejection occurs with conditional
probability at least $1/2$. Hence
$$
\mathbb E_s[d_cD_I]
\geq d_c(1-Q_I)\geq\frac{d_c}{2}.
$$

\xhdr{Step 2: Bound underpricing when acceptance is likely}
Suppose $Q_I>1/2$ and $n\geq T/3$.
For $k=0,\ldots,L-1$, let $\bar p_k$ be the price in period $s+k$
along the counterfactual path on which all preceding prices in the
block are accepted. Define
$$
q_k:=q(n+k,\bar p_k),
\qquad
G_k:=R_{n+k}-q_k.
$$
The prices and thresholds are fixed by the initial public history,
the seed $U$, and the assumed acceptances.
On $C_I$, this path agrees with the realized interaction, so
$p_{s+k}=\bar p_k$, $K_{s+k-1}=n+k$, and $G_k\geq0$ by \Cref{lem:lower-threshold-slack}.

The increment $\Delta_k:=R_{n+k}-R_n$ has conditional distribution
$\operatorname{Binomial}(k,1/2)$.
For $k\geq256$, set $w_k:=\lfloor\sqrt{k}/16\rfloor$.
The binomial point-mass bound\footnote{For
$X\sim\operatorname{Binomial}(k,1/2)$, Stirling's formula gives
$\sup_m\mathbb P(X=m)\leq k^{-1/2}$ for $k\geq1$.}
implies
$$
\mathbb P_s(0\leq G_k<w_k)
\leq\frac{w_k}{\sqrt{k}}\leq\frac1{16},
$$
because this event restricts $\Delta_k$ to at most $w_k$ integer values.
Since $G_k\geq0$ on $C_I$,
\begin{equation}
\label{eq:lower-block-gap-probability}
\mathbb P_s(C_I\cap\{G_k\geq w_k\})
\geq Q_I-\frac1{16}>\frac7{16}.
\end{equation}

On $\mathcal E_T$ from \Cref{lem:lower-central-path},
we have $T/3\leq n+k\leq T$ and
$|R_{n+k}/(n+k)-1/2|\leq\eta$.
For sufficiently large $T$, also $n+k\geq n_F$ and
$w_k/(n+k)\leq\eta$.
Thus \Cref{lem:lower-threshold-slack} implies that, on
$C_I\cap\{G_k\geq w_k\}\cap\mathcal E_T$,
\begin{equation}
\label{eq:lower-block-price-gap}
\mu(n+k,R_{n+k})-\bar p_k
\geq\frac{\lambda_Fw_k}{n+k}
\geq\frac{\lambda_Fw_k}{T}.
\end{equation}

\xhdr{Step 3: Sum and average the conditional bounds}
Sum over $k=\lceil L/2\rceil,\ldots,L-1$, so that $k\geq256$.
Since $w_k\geq\sqrt{k}/32$, there is a universal constant
$\gamma>0$ such that
$$
\gamma L^{3/2}\leq\sum_k w_k\leq L^{3/2}\leq T.
$$
Using
\eqref{eq:lower-block-gap-probability} and \eqref{eq:lower-block-price-gap},
and subtracting at most $\mathbb P_s(\mathcal E_T^c)$ from the event
probability, gives
$$
\begin{aligned}
\mathbb E_s[W_I]
&\geq\frac{7\lambda_F}{16T}\sum_k w_k
-\frac{\lambda_F}{T}\sum_k w_k\,
  \mathbb P_s(\mathcal E_T^c)\\
&\geq b_F-\lambda_F\mathbb P_s(\mathcal E_T^c)
\end{aligned}
$$
for some constant $b_F>0$, because $L\geq T^{2/3}/2$ for large $T$.

Set $b:=\min\{d_c/2,b_F\}$.
Combining the two cases, and using nonnegativity of losses when
$n<T/3$, yields
$$
\mathbb E_s[W_I+d_cD_I]
\geq b\,\mathbf 1\{n\geq T/3\}
-\lambda_F\mathbb P_s(\mathcal E_T^c).
$$
Averaging over the block-start history and the policy's randomization
proves the lemma.
\end{proof}

We now combine
\Cref{lem:lower-regret-decomposition,lem:lower-block-tradeoff,lem:lower-central-path}
to prove \Cref{thm:lower-bound}.

\begin{proof}[Proof of \Cref{thm:lower-bound}]
Fix an arbitrary admissible, possibly randomized, policy
$A\in\mathcal A$ and set
$\theta=1/2$. Maintain \Cref{ass:regular-general-prior} and
$c\in[0,\varepsilon]$. The decomposition
\eqref{eq:lower-regret-decomposition} in
\Cref{lem:lower-regret-decomposition} gives
$$
\operatorname{Reg}_T\left(A;\frac12\right)
\geq
\mathbb E_{Y,U}[W_T+d_cD_T]-B,
\qquad d_c:=\frac{1/2-c}{2}>0.
$$

We complete the proof in two steps.

\xhdr{Step 1: Bound regret with many rejections}
We first show that rejections alone yield the desired lower bound if at least $T/4$ rejections occur with sufficiently large probability.

Define the event $H:=\{D_T\geq T/4\}$ and let $h:=\mathbb P_{Y,U}(H)$.
Since $D_T\geq T/4$ on $H$ and $D_T\geq0$ otherwise,
$\mathbb E_{Y,U}[D_T]\geq (T/4)h$.

If $h\geq T^{-2/3}$, then \Cref{lem:lower-regret-decomposition},
together with the nonnegativity of the underpricing loss $W_T$, gives
$$
\begin{aligned}
\operatorname{Reg}_T\left(A;\frac12\right)
&\geq d_c\mathbb E_{Y,U}[D_T]-B\\
&\geq \frac{d_cT}{4}h-B\\
&\geq \frac{d_c}{4}T^{1/3}-B.
\end{aligned}
$$
Because $d_c>0$ and $B$ is independent of $T$ and $A$, this establishes
the desired $\Omega(T^{1/3})$ lower bound for sufficiently large $T$.
It therefore remains to consider policies with $h<T^{-2/3}$.

\xhdr{Step 2: Aggregate the late blocks}
Suppose that $h<T^{-2/3}$. 
Set $L:=\lfloor T^{2/3}\rfloor$ and $M:=\lfloor T/(4L)\rfloor$.
We partition the final $ML$ periods into the disjoint blocks
$$
I_r:=\{T-ML+rL+1,\ldots,T-ML+(r+1)L\},
$$
for $r=0,\ldots,M-1$.
Since $ML\leq T/4$, every selected block lies in the last quarter of the horizon.

To apply \Cref{lem:lower-block-tradeoff}, we first verify that sufficiently many calls have occurred before each block.
On $H^c$, fewer than $T/4$ periods are rejected over the entire horizon. 
A block starting at period $s$ therefore satisfies
$$
K_{s-1}\geq s-1-D_T
\geq T-ML-D_T>T/2.
$$
Thus $\mathbb P_{Y,U}(K_{s-1}\geq T/3)\geq1-h$ for every selected
block. Choosing late blocks ensures that the required initial
call count is reached with high probability.

The remaining error term comes from the event $\mathcal E_T^c$,
whose probability is bounded in \Cref{lem:lower-central-path}.
For sufficiently large $T$, \eqref{eq:lower-central-probability}
gives $\lambda_F\mathbb P_Y(\mathcal E_T^c)\leq b/4$, while
$h<T^{-2/3}\leq1/4$.
Applying the blockwise bound \eqref{eq:lower-block-loss} therefore yields
\begin{equation}
\label{eq:lower-selected-block-loss}
\begin{aligned}
\mathbb E_{Y,U}[W_{I_r}+d_cD_{I_r}]
&\geq b(1-h)-\lambda_F\mathbb P_Y(\mathcal E_T^c)\\
&\geq \frac b2
\end{aligned}
\end{equation}
for every $r$. Hence each block contributes a constant expected
loss through rejections, underpricing, or both.

Since the blocks are disjoint and each period contributes
nonnegatively to $W_T$ and $D_T$, we can sum their losses to obtain
$$
W_T+d_cD_T
\geq
\sum_{r=0}^{M-1}(W_{I_r}+d_cD_{I_r}).
$$
For sufficiently large $T$, the number of selected blocks satisfies
$M\geq T^{1/3}/8$.
Taking expectations and applying
\eqref{eq:lower-regret-decomposition} and
\eqref{eq:lower-selected-block-loss}, we obtain
$$
\operatorname{Reg}_T\left(A;\frac12\right)
\geq\frac{bM}{2}-B
\geq\frac{b}{16}T^{1/3}-B.
$$
Since $b>0$ and $B$ is independent of $T$ and $A$, this establishes
the desired $\Omega(T^{1/3})$ lower bound in the remaining case.

Combining the two cases, take
$\kappa:=\min\{d_c/4,b/16\}>0$ and a finite integer $T_0$ such that
$\operatorname{Reg}_T(A;1/2)\geq\kappa T^{1/3}-B$ for all $T\geq T_0$.
Regret is nonnegative at every horizon, so choosing
$K:=B+\kappa T_0^{1/3}>0$ covers all integers $T\geq2$.
Both constants depend only on $(F,c,\varepsilon)$ and not on the policy.
Because $1/2\in\Theta_\varepsilon$, taking the infimum over
$A\in\mathcal A$ gives the statement of \Cref{thm:lower-bound}.
\end{proof}

\section{Omitted Proofs in \texorpdfstring{\Cref{sec:big-cost}}{Section \ref*{sec:big-cost}}}

\label{apx:big cost}

\subsection{Proof of the Upper Bound}
\label{apx:pf-big-cost-upper}

Throughout this appendix, assume \Cref{ass:regular-general-prior} and fix
$c\in(\varepsilon,1-\varepsilon)$.
We first prove that CAT-R in \Cref{alg:cst-big-cost} has regret
$O(\sqrt T)$ uniformly over $\theta\in\Theta_\varepsilon$.
The proof bounds three sources of loss: underpricing, calibration rejections, and incorrect service-continuation decisions.
Posterior concentration then relates these losses to the dynamic full-information benchmark.

Use the accepted-call notation from \Cref{apx:pf-lower}.
For a fixed type $\theta$, let $Y_1,Y_2,\ldots$ be independent
Bernoulli$(\theta)$ potential outcomes, and set
$R_n:=\sum_{i=1}^nY_i$ and $K_t:=\sum_{s=1}^t a_s$, with $R_0=K_0=0$.
Each accepted call consumes the next outcome.
Thus, before period $t$, the user's state is
$(\callCount{t},\successCount{t})=(K_{t-1},R_{K_{t-1}})$.
Write $\widehat\theta_n:=R_n/n$ for $n\geq1$ and
$(x)_+:=\max\{x,0\}$.

The sensitivity bound in \Cref{lem:posterior-upper-sensitivity} already
bounds how prices change with the success count.
The next lemma bounds the difference between the user's posterior
mean and the true type, which is needed because the benchmark can choose
when to serve.

\begin{lemma}[Posterior concentration]
\label{lem:catr-posterior}
There are finite constants $C_p,C_\mu>0$, depending only on
$(F,\varepsilon)$, such that for all integers $n\geq1$ and
$0\leq s\leq n$,
$$
\left|\mu(n,s)-
\min\{1-\varepsilon,\max\{\varepsilon,s/n\}\}\right|
\leq\frac{C_p}{\sqrt n}.
$$
Moreover, for every $\theta\in\Theta_\varepsilon$ and integer $T\geq1$,
$$
\sum_{n=0}^{T-1}\mathbb E_\theta|\mu(n,R_n)-\theta|
\leq C_\mu\sqrt T.
$$
\end{lemma}

\begin{proof}[Proof of \Cref{lem:catr-posterior}]
The proof proceeds in two steps.

\xhdr{Step 1: Bound the posterior second moment}
We first compare the posterior with the normalized likelihood, whose
curvature gives a bound valid even at the support endpoints.

Fix $(n,s)$ and let $g$ be the probability density on
$\Theta_\varepsilon$ proportional to $u^s(1-u)^{n-s}$.
Let $V$ have density $g$.
Put $\ell(u):=s\log u+(n-s)\log(1-u)$,
$z:=\min\{1-\varepsilon,\max\{\varepsilon,s/n\}\}$, and
$k:=n/(1-\varepsilon)^2$.
Then $\ell''(u)\leq-k$ throughout the interval.
The maximizing point $z$ satisfies
$(u-z)\ell'(z)\leq0$, including when $z$ is an endpoint, so
$$
(u-z)\ell'(u)\leq-k(u-z)^2.
$$
Since $g'=\ell'g$, integration by parts gives
$$
\mathbb E_g[(V-z)\ell'(V)]
=[(u-z)g(u)]_{\varepsilon}^{1-\varepsilon}-1\geq-1.
$$
The boundary term is nonnegative because $z\in\Theta_\varepsilon$.
Combining these inequalities yields
$\mathbb E_g[(V-z)^2]\leq1/k$.

The posterior density is proportional to $f(u)g(u)$.
By \Cref{ass:regular-general-prior}, its ratio to $g(u)$ is at most
$\overline f/\underline f$ almost everywhere.
Its second moment about $z$ is therefore at most
$\overline f/(\underline f k)$.
Cauchy--Schwarz proves the first claim with
$C_p:=(1-\varepsilon)\sqrt{\overline f/\underline f}$.

\xhdr{Step 2: Sum the deviations from the true type}
Projection onto $\Theta_\varepsilon$ cannot increase the distance to
$\theta\in\Theta_\varepsilon$. Hence
$$
|\mu(n,R_n)-\theta|
\leq\frac{C_p}{\sqrt n}+|\widehat\theta_n-\theta|.
$$
Since $\operatorname{Var}_\theta(\widehat\theta_n)
=\theta(1-\theta)/n\leq1/(4n)$, the expected second term is at most
$1/(2\sqrt n)$.
Finally, $|\mu(0,0)-\theta|\leq1$ and
$\sum_{n=1}^{T-1}n^{-1/2}\leq2\sqrt T$.
Taking $C_\mu:=2C_p+2$ proves the sum bound.
\end{proof}

We next verify the public count bounds maintained by CAT-R.
During calibration, the threshold variable $b$ increases while the stored
variable $L$ remains unchanged; the proof therefore uses $b$ as the current
lower bound in that phase.

\begin{lemma}[Count and pricing bounds]
\label{lem:catr-count-bounds}
CAT-R is admissible. 
On every outcome sequence, its public upper bound $U$ satisfies $0\leq\frac{U-R_n}{n}\leq\frac{2}{\sqrt n}$ whenever the retirement test is evaluated after $n\geq1$ calls.

For horizon $T\geq1$, let $J_T$ count rejected queries before
retirement and define the total underpricing by
$$
W_T:=\sum_{t=1}^T
a_t\bigl(\mu(K_{t-1},R_{K_{t-1}})-p_t\bigr).
$$
Then, on every outcome sequence,
$$
W_T\leq4\Lambda_{\varepsilon,F}\sqrt T,
\qquad
J_T\leq(1+\sqrt2)\sqrt T,
$$
where $\Lambda_{\varepsilon,F}$ is the constant in
\Cref{lem:posterior-upper-sensitivity}.
\end{lemma}

\begin{proof}[Proof of \Cref{lem:catr-count-bounds}]
The proof proceeds in three steps.

\xhdr{Step 1: Verify the count bounds and query decisions}
Consider a cycle starting at the exact state $(n_0,s_0)$, and set $E:=\lceil\sqrt{n_0+1}\rceil$.
Define $\ell=s_0$ during tracking and $\ell=b$ during calibration.
We verify that
$$
0\leq\ell\leq R_n\leq U\leq n,
\qquad U-\ell\leq E
$$
throughout the cycle, before any retirement test or posted price.
These bounds hold at the start of the cycle, where
$\ell=R_{n_0}=U=s_0$.
By \Cref{lem:posterior-upper-sensitivity}, $\mu(n,s)$ is strictly increasing
in the integer success count; the case $n=0$ has only the state $(0,0)$.
Purchase at equality therefore implies that a query at $\mu(n,q)$ is
accepted exactly when $q\leq R_n$.

During tracking, the query is $q=s_0\leq R_n$.
After each call, $n$ and $U$ increase by one, while $R_n$ increases by either
zero or one. Keeping $\ell=s_0$ preserves the bounds, and their width is at
most the number of tracking calls, hence at most $E$.

If calibration is reached, after $k$ accepted calibration queries we have
$b=s_0+k$ and $n=n_0+E+k$.
Since $s_0\leq n_0$ and $E\geq1$, the next threshold $b+1$ is at most $n$.
An acceptance implies that the current success count is at least $b+1$.
Increasing $b,n,U$ by one then preserves both the count bounds and their
width. A rejection implies $R_n<b+1$; together with $R_n\geq b$ and
integrality, this gives $R_n=b$.
The update $(L,U)=(b,b)$ makes the next reset exact.
Stopping or retiring can only end this sequence of valid updates.

All prices and state variables depend only on public purchase decisions and
the known prior, cost, and horizon.
Every queried posterior mean lies in $[\varepsilon,1-\varepsilon]$, and
retirement posts the feasible price $1$, which is always rejected.
On an unreachable public history, define the policy to post $1$ thereafter. This establishes admissibility.

\xhdr{Step 2: Bound underpricing on accepted calls}
The first price is $\mu(0,0)$, so its underpricing is zero.
For every later accepted call, the query is at $\ell$ or $\ell+1$, and the
success-count difference is at most $E$.
Since $n_0\leq n$ and $n\geq1$,
$$
E\leq\lceil\sqrt{n+1}\rceil\leq\sqrt n+1\leq2\sqrt n.
$$
For the middle inequality, if $j=\lceil\sqrt{n+1}\rceil$, then
$(j-1)^2<n+1$, so integrality gives $(j-1)^2\leq n$.
Summing the sensitivity bound over the intervening states gives
$$
\mu(n,R_n)-p_t\leq\frac{\Lambda_{\varepsilon,F}E}{n}
\leq\frac{2\Lambda_{\varepsilon,F}}{\sqrt n}.
$$
Each call count occurs before at most one accepted call.
Summing over $1\leq n<K_T\leq T$ proves the bound on $W_T$.
The count bounds also give
$0\leq U-R_n\leq U-\ell\leq E\leq2\sqrt n$ before every retirement test with $n\geq1$.
This proves the stated count-error bound, including when the test triggers retirement.

\xhdr{Step 3: Count calibration rejections}
Every active rejection ends one cycle, and each completed cycle includes
its full tracking phase.
Let $n_0=0,n_1,\ldots,n_{J_T}$ be the call counts after successive completed
cycles. Then $n_{j+1}\geq n_j+\sqrt{n_j+1}$.
Writing $x=n_j+1\geq1$ gives
$$
\begin{aligned}
\sqrt{n_{j+1}+1}-\sqrt{n_j+1}
&\geq\sqrt{x+\sqrt x}-\sqrt x\\
&=\frac{1}{\sqrt{1+x^{-1/2}}+1}\geq\frac{1}{1+\sqrt2}.
\end{aligned}
$$
Summing over completed cycles and using $n_{J_T}\leq K_T\leq T$ yields
$$
J_T\leq(1+\sqrt2)(\sqrt{K_T+1}-1)
\leq(1+\sqrt2)\sqrt T.
$$
An unfinished final cycle contributes no additional active rejection.
\end{proof}

The remaining task is to bound losses from service decisions.
For types below cost, the sum of the retirement bonuses limits continued
service. For types above cost, an expected maximal deviation limits the
loss from retiring too early.

\begin{lemma}[Retirement bonus bounds]
\label{lem:catr-bonus}
For any integer $T\geq1$ and $\theta\in\Theta_\varepsilon$, let
$r_T(n):=\sqrt{\log(T/n)/n}$ for $1\leq n\leq T$, and define
$$
Z_T^{\mathrm{ret}}
:=\max_{1\leq n\leq T}
(\theta-\widehat\theta_n-r_T(n))_+.
$$
The following bounds hold with a universal constant $C_z<\infty$:
$$
\sum_{n=1}^{T-1}r_T(n)\leq\sqrt{2\pi T},
\qquad
\mathbb E_\theta Z_T^{\mathrm{ret}}\leq\frac{C_z}{\sqrt T}.
$$
\end{lemma}

\begin{proof}[Proof of \Cref{lem:catr-bonus}]
The proof proceeds in two steps.

\xhdr{Step 1: Sum the bonuses}
The bonus is decreasing on $(0,T]$.
An integral comparison, followed by $x=Te^{-u}$, gives
$$
\begin{aligned}
\sum_{n=1}^{T-1}r_T(n)
&\leq\int_0^T\sqrt{\frac{\log(T/x)}x}\,\mathrm dx\\
&=\sqrt T\int_0^\infty u^{1/2}e^{-u/2}\,\mathrm du=\sqrt{2\pi T}.
\end{aligned}
$$

\xhdr{Step 2: Bound the largest downward error}
We first obtain a maximal concentration bound for
$M_n:=n\theta-R_n$.
For any $\lambda\geq0$, Hoeffding's lemma applied to the independent,
mean-zero variables $\theta-Y_i$, each in an interval of length one, shows that
$\exp(\lambda M_n-n\lambda^2/8)$ is a nonnegative supermartingale.
Stopping it at the first crossing of $a\geq0$, capped at an integer $m$,
and taking $\lambda=4a/m$ gives
$$
\mathbb P_\theta\!\left(\max_{1\leq n\leq m}M_n\geq a\right)
\leq\exp(-2a^2/m).
$$
Indeed, on a crossing the stopped exponential is at least
$\exp(\lambda a-m\lambda^2/8)$, whereas its expectation is at most one
by bounded optional stopping.

Partition the call counts into the nonempty blocks
$B_j:=\{n\in\mathbb N:T/2^{j+1}<n\leq T/2^j\}$, for $j\geq0$,
and write $m_j:=T/2^{j+1}$.
If $\theta-\widehat\theta_n-r_T(n)>x$ for some $n\in B_j$ and $x\geq0$,
then
$$
M_n>nx+\sqrt{n\log(T/n)}
\geq m_jx+\sqrt{m_jj\log2}.
$$
All indices in the block are at most $\lfloor2m_j\rfloor$.
The maximal bound therefore gives
$$
\begin{aligned}
&\mathbb P_\theta\!\left(
\max_{n\in B_j}(\theta-\widehat\theta_n-r_T(n))>x
\right)\\
&\qquad\leq
\exp\!\left[-(\sqrt{m_j}x+\sqrt{j\log2})^2\right]
\leq2^{-j}e^{-m_jx^2}.
\end{aligned}
$$
A union bound and integration of the tail probabilities now yield
$$
\mathbb E_\theta Z_T^{\mathrm{ret}}
\leq\sum_{j\geq0}2^{-j}\int_0^\infty e^{-m_jx^2}\,\mathrm dx
=\frac{\sqrt{\pi/2}}{\sqrt T}\sum_{j\geq0}2^{-j/2}.
$$
Empty blocks may be included in this upper bound.
The geometric series is finite, proving the claim with
$C_z=\sqrt{\pi/2}/(1-2^{-1/2})$.
\end{proof}

We can now bound the cost of serving too often or too rarely.
The following statement is uniform even when $\theta$ is arbitrarily close
to $c$.

\begin{lemma}[Loss from service decisions]
\label{lem:catr-service-loss}
There is a universal constant $C_s<\infty$ such that CAT-R satisfies
$$
\begin{aligned}
(\theta-c)\mathbb E_\theta[T-K_T]&\leq C_s\sqrt T,
&&\theta\geq c,\\
(c-\theta)\mathbb E_\theta[K_T]&\leq C_s\sqrt T,
&&\theta<c.
\end{aligned}
$$
\end{lemma}

\begin{proof}[Proof of \Cref{lem:catr-service-loss}]
The proof proceeds in two steps.

\xhdr{Step 1: Bound missed service above cost}
Let $H_T$ count all periods assigned price $1$ after retirement, including
the period in which the retirement test first succeeds.
Every period is either an accepted call, an active rejection, or a retired
period, so $T=K_T+J_T+H_T$.
Put $\Delta:=\theta-c\geq0$.
If retirement occurs after $n$ calls, the test and the count bound give
$$
\widehat\theta_n+r_T(n)\leq U/n+r_T(n)<c.
$$
Consequently, $\Delta H_T\leq T Z_T^{\mathrm{ret}}$.
This inequality also holds when retirement never occurs or $\Delta=0$.
Since $\Delta\leq1$, \Cref{lem:catr-count-bounds,lem:catr-bonus} imply
$$
\Delta\mathbb E_\theta[T-K_T]
\leq\mathbb E_\theta[J_T]+T\mathbb E_\theta Z_T^{\mathrm{ret}}
\leq(1+\sqrt2+C_z)\sqrt T.
$$

\xhdr{Step 2: Bound continued service below cost}
Put $d:=c-\theta>0$.
On the event $\{K_T>n\}$ for $1\leq n\leq T-1$, consider the period of the
$(n+1)$-st accepted call.
Its retirement test did not trigger, and
\Cref{lem:catr-count-bounds} gives $(U-R_n)/n\leq2/\sqrt n$.
Thus
$$
d\leq\widehat\theta_n-\theta+\frac2{\sqrt n}+r_T(n).
$$
Multiplying by $\mathbf1_{\{K_T>n\}}$ and taking expectations yields
$$
d\,\mathbb P_\theta(K_T>n)
\leq\mathbb E_\theta(\widehat\theta_n-\theta)_+
     +\frac2{\sqrt n}+r_T(n)
\leq\frac{5}{2\sqrt n}+r_T(n).
$$
This uses an unconditional bound at the deterministic call count $n$;
no conditional unbiasedness is required.
The first call contributes at most $d\leq1$.
Using $\mathbb E_\theta K_T=\sum_{n=0}^{T-1}\mathbb P_\theta(K_T>n)$
and \Cref{lem:catr-bonus}, we obtain
$$
d\,\mathbb E_\theta K_T
\leq1+5\sqrt T+\sqrt{2\pi T}.
$$
The two cases establish the claim.
\end{proof}

The preceding lemmas bound CAT-R's losses on the potential-outcome
sequence. 
We now combine these bounds to prove the upper bound.

\begin{proof}[Upper-bound part of \Cref{thm:big-cost-minimax}]
The proof proceeds in two steps.

\xhdr{Step 1: Compare profits on the same outcome sequence}
Fix $\theta\in\Theta_\varepsilon$.
An optimal benchmark policy exists by backward induction over the finitely
many reachable states in the Bellman recursion.
Whenever it serves, it can charge $\mu(n,R_n)$.
If it makes $M\leq T$ calls, its realized profit is
$\sum_{n=0}^{M-1}(\mu(n,R_n)-c)$.
For every outcome sequence and every such $M$,
$$
\sum_{n=0}^{M-1}(\mu(n,R_n)-c)
\leq T(\theta-c)_+
+    \sum_{n=0}^{T-1}|\mu(n,R_n)-\theta|.
$$
Hence \Cref{lem:catr-posterior} implies
$V_T^{\mathrm{OFF}}(\theta)\leq T(\theta-c)_++C_\mu\sqrt T$.
This comparison allows both waiting and calls with negative current margins.

CAT-R's profit satisfies the pathwise identity
$$
\sum_{t=1}^T a_t(p_t-c)
=(\theta-c)K_T+
  \sum_{n=0}^{K_T-1}(\mu(n,R_n)-\theta)-W_T.
$$
Bounding the partial sum below by
$-\sum_{n=0}^{T-1}|\mu(n,R_n)-\theta|$ and applying
\Cref{lem:catr-posterior,lem:catr-count-bounds} gives
$$
\Pi_T(\CATR;\theta)
\geq(\theta-c)\mathbb E_\theta K_T
-(C_\mu+4\Lambda_{\varepsilon,F})\sqrt T.
$$
Both comparisons hold for outcome-dependent call counts, including when
the horizon ends within calibration.

\xhdr{Step 2: Combine pricing and service-decision losses}
Subtracting the last bound from the benchmark bound yields
$$
\begin{aligned}
\operatorname{Reg}_T(\CATR;\theta)
\leq{}&T(\theta-c)_+-(\theta-c)\mathbb E_\theta K_T\\
&+(2C_\mu+4\Lambda_{\varepsilon,F})\sqrt T.
\end{aligned}
$$
The first line equals $(\theta-c)\mathbb E_\theta(T-K_T)$ when
$\theta\geq c$, and $(c-\theta)\mathbb E_\theta K_T$ when $\theta<c$.
\Cref{lem:catr-service-loss} therefore bounds it by $C_s\sqrt T$.
For $C:=2C_\mu+4\Lambda_{\varepsilon,F}+C_s$, we conclude
$$
\mathcal R_T^*(\Theta_\varepsilon)
\leq\sup_{\theta\in\Theta_\varepsilon}
\operatorname{Reg}_T(\CATR;\theta)
\leq C\sqrt T.
$$
The constant depends only on $(F,\varepsilon)$.
All estimates include $\theta=c$, the support endpoints, and $T=1$.
\end{proof}

\subsection{Proof of the Lower Bound}
\label{apx:pf-big-cost-lower}

For the lower bound, the hard types lie on opposite sides of the serving cost. 
Serving too often is costly under the lower type, while serving too rarely gives up profit under the higher type.
We first account for deviations of the user's posterior mean from the true type, then relate regret to these two call-count errors.
An information bound shows that no policy can make both errors small.

Fix an arbitrary admissible, possibly randomized policy $A$.
Write $n:=K_T$ for its total number of calls, as in the main-text sketch,
and use $j$ for deterministic call counts below.
Set
$$
d_c:=\min\{c-\varepsilon,1-\varepsilon-c\}>0,
\qquad
\theta_-:=c-\delta,\quad\theta_+:=c+\delta,
$$
where $0<\delta\leq d_c$ will be chosen at the end.
Both types belong to $\Theta_\varepsilon$.
All constants in this subsection are independent of $A$ and $T$.

We first bound the total deviations that exceed a fixed fraction of the distance $\delta$ from cost.

\begin{lemma}[Summable posterior deviations]
\label{lem:cost-excursions}
Set $D_F:=4C_p^2+3$, using $C_p$ from \Cref{lem:catr-posterior}.
For every $\theta\in\Theta_\varepsilon$ and $0<\delta\leq1$,
$$
\begin{aligned}
\sum_{j=0}^{\infty}
\mathbb E_\theta[(\mu(j,R_j)-\theta-\delta/2)_+]
&\leq\frac{D_F}{\delta},\\
\sum_{j=0}^{\infty}
\mathbb E_\theta[(\theta-\mu(j,R_j)-\delta/2)_+]
&\leq\frac{D_F}{\delta}.
\end{aligned}
$$
\end{lemma}

\begin{proof}[Proof of \Cref{lem:cost-excursions}]
We prove the upper-tail bound; the lower-tail proof reverses the signs.
For $a>0$ and $j\geq1$, the posterior approximation in
\Cref{lem:catr-posterior} gives
$$
(\mu(j,R_j)-\theta-a)_+
\leq(\widehat\theta_j-\theta-a/2)_+
   +(C_p/\sqrt j-a/2)_+.
$$
Here projection onto $\Theta_\varepsilon$ cannot increase a positive
deviation above $\theta\in\Theta_\varepsilon$, or a negative deviation below it.
Hoeffding's inequality, Tonelli's theorem, and
$e^{2x^2}-1\geq2x^2$ yield
$$
\begin{aligned}
\sum_{j=1}^{\infty}
\mathbb E_\theta[(\widehat\theta_j-\theta-a/2)_+]
&\leq\int_{a/2}^{\infty}\sum_{j=1}^{\infty}e^{-2jx^2}\,\mathrm dx\\
&=\int_{a/2}^{\infty}\frac{\mathrm dx}{e^{2x^2}-1}
\leq\frac1a.
\end{aligned}
$$
The deterministic part is decreasing in $j$, so
$$
\sum_{j=1}^{\infty}(C_p/\sqrt j-a/2)_+
\leq\int_0^\infty(C_p/\sqrt x-a/2)_+\,\mathrm dx
=\frac{2C_p^2}{a}.
$$
The $j=0$ contribution is at most one.
Taking $a=\delta/2$ gives
$1+(4C_p^2+2)/\delta\leq D_F/\delta$.
The lower-tail bound follows from the lower-tail version of Hoeffding's
inequality and the same calculation.
\end{proof}

These deviation bounds let us compare service counts without assuming
that early profit margins already have the sign of $\theta-c$.

\begin{lemma}[Regret and the number of calls]
\label{lem:cost-call-counts}
For the two types $\theta_\pm=c\pm\delta$ defined above, every admissible
policy satisfies
$$
\begin{aligned}
\operatorname{Reg}_T(A;\theta_-)
&\geq\frac{\delta}{2}\mathbb E_{A,\theta_-}[n]
       -\frac{D_F}{\delta},\\
\operatorname{Reg}_T(A;\theta_+)
&\geq\frac{\delta}{2}\mathbb E_{A,\theta_+}[T-n]
       -\frac{D_F}{\delta}.
\end{aligned}
$$
\end{lemma}

\begin{proof}[Proof of \Cref{lem:cost-call-counts}]
Use the same potential-outcome sequence for the online policy and each
benchmark comparison. Since an accepted price cannot exceed the user's
posterior mean, pathwise
$$
\sum_{t=1}^T a_t(p_t-c)
\leq\sum_{j=0}^{n-1}(\mu(j,R_j)-c).
$$
At $\theta_-$, immediate retirement is feasible for the benchmark and yields
zero. Moreover,
$$
\mu(j,R_j)-c
\leq-\delta/2+(\mu(j,R_j)-\theta_- -\delta/2)_+.
$$
Sum over $j<n$, bound the positive terms by their sum over all $j\geq0$,
and apply \Cref{lem:cost-excursions} to obtain the first inequality.

At $\theta_+$, serving in every period at the posterior mean is a feasible
benchmark policy.
Subtracting the online prefix bound from this policy's profit leaves
$\sum_{j=n}^{T-1}(\mu(j,R_j)-c)$.
Each term is at least
$\delta/2-(\theta_+-\mu(j,R_j)-\delta/2)_+$.
The second inequality follows from \Cref{lem:cost-excursions}.
The comparisons hold on every outcome sequence, so they do not
require the random total $n$ to be independent of the outcomes.
\end{proof}

To keep both call-count errors small, a policy would need to distinguish
the two types from purchase decisions.
The next lemma bounds how much information those observations contain.

\begin{lemma}[Information in the provider's observations]
\label{lem:cost-public-kl}
Let $P_-^A$ and $P_+^A$ be the laws under $\theta_-$ and $\theta_+$ of
the public history $\pubHis{T+1}=(p_1,a_1,\ldots,p_T,a_T)$.
Then
$$
\operatorname{KL}(P_-^A\|P_+^A)
\leq L_\varepsilon\delta^2\mathbb E_{A,\theta_-}[n],
\qquad
L_\varepsilon:=\frac{4}{\varepsilon(1-\varepsilon)}.
$$
\end{lemma}

\begin{proof}[Proof of \Cref{lem:cost-public-kl}]
Augment the public history with the policy's random seed, independent of
the type and outcomes, and the outcomes of accepted calls.
Conditional on this augmented history, the same pricing policy and posterior
map determine the next price and purchase decision under either type.
An accepted call adds a Bernoulli outcome; a rejected period adds no outcome.
The chain rule for relative entropy therefore gives
$$
\operatorname{KL}(Q_-^A\|Q_+^A)
=\mathbb E_{A,\theta_-}[n]\,
\operatorname{kl}(\theta_-,\theta_+),
$$
where $Q_\pm^A$ are the augmented-history laws and
$\operatorname{kl}$ is Bernoulli relative entropy.
The common seed contributes zero relative entropy.
Discarding private outcomes and the seed cannot increase relative entropy,
so the same expression bounds the public-history divergence.
Finally,
$\operatorname{kl}(u,v)\leq(u-v)^2/[v(1-v)]$ for $u,v\in(0,1)$.
Since $\theta_\pm\in[\varepsilon,1-\varepsilon]$ and
$\theta_+-\theta_-=2\delta$, this proves the claim.
\end{proof}

The last step combines the call-count and information bounds.
The constant in $\delta\asymp T^{-1/2}$ must account for the
$D_F/\delta$ posterior correction, so we choose it explicitly.

\begin{proof}[Lower-bound part of \Cref{thm:big-cost-minimax}]
The proof proceeds in three steps. We first turn small regret at both
types into a test that distinguishes them, then bound the accuracy of
this test, and finally choose the separation between the types.

\xhdr{Step 1: Relate small regret to opposite service decisions}
Fix an arbitrary admissible policy $A$ and write $n:=K_T$.
For now, fix $0<\delta\leq d_c$ and recall that
$\theta_\pm=c\pm\delta$.
Small regret requires few calls under $\theta_-$ but many calls
under $\theta_+$.
To quantify these requirements, let $r$ be the larger regret at the
two types and define $u$ to include the posterior-deviation correction:
$$
\begin{aligned}
r&:=\max\{\operatorname{Reg}_T(A;\theta_-),
          \operatorname{Reg}_T(A;\theta_+)\},\\
u&:=D_F+\delta r.
\end{aligned}
$$
Rearranging the two inequalities in \Cref{lem:cost-call-counts} gives
$$
\mathbb E_{A,\theta_-}[n]\leq\frac{2u}{\delta^2},
\qquad
\mathbb E_{A,\theta_+}[T-n]\leq\frac{2u}{\delta^2}.
$$
These inequalities also imply $u\geq0$, since both expected
call-count errors are nonnegative.

The total call count is observable from the public history.
Thus the event $B:=\{n\geq T/2\}$ defines a test: choose $\theta_+$
on $B$ and $\theta_-$ on $B^c$.
An error under $\theta_-$ requires at least $T/2$ calls, whereas
an error under $\theta_+$ leaves more than $T/2$ periods unserved.
Applying Markov's inequality to each error and adding the bounds yields
$$
P_-^A(B)+P_+^A(B^c)\leq\frac{8u}{T\delta^2}.
$$

\xhdr{Step 2: Bound the testing error from below}
We next show that the information available to the provider prevents
this test from being too accurate.
For any probability laws $P,Q$ and event $B$, the
Bretagnolle--Huber inequality states
$$
P(B)+Q(B^c)\geq\frac12
\exp\{-\operatorname{KL}(P\|Q)\}.
$$
For completeness, suppose first that $\operatorname{KL}(P\|Q)<\infty$.
Let $h:=\int\sqrt{\mathrm dP\,\mathrm dQ}$ denote their Hellinger
affinity.
Jensen's inequality gives
$h\geq e^{-\operatorname{KL}(P\|Q)/2}$, while Cauchy--Schwarz bounds
their total variation distance by $\sqrt{1-h^2}$.
Consequently,
$$
\begin{aligned}
P(B)+Q(B^c)
&\geq1-\operatorname{TV}(P,Q)\\
&\geq1-\sqrt{1-h^2}\\
&\geq\frac{h^2}{2}
\geq\frac12e^{-\operatorname{KL}(P\|Q)}.
\end{aligned}
$$
If the relative entropy is infinite, the right-hand side is zero,
so the inequality still holds.

To apply this inequality, \Cref{lem:cost-public-kl} and the
call-count bound from Step 1 give
$$
\operatorname{KL}(P_-^A\|P_+^A)
\leq L_\varepsilon\delta^2\mathbb E_{A,\theta_-}[n]
\leq2L_\varepsilon u.
$$
Combining the resulting lower bound with the upper bound on the same
testing error yields
$$
\frac{8u}{T\delta^2}
\geq P_-^A(B)+P_+^A(B^c)
\geq\frac12e^{-2L_\varepsilon u}.
$$
Equivalently, every admissible policy must satisfy
\begin{equation}
\label{eq:catr-lower-testing}
T\delta^2\leq16u\exp(2L_\varepsilon u).
\end{equation}

\xhdr{Step 3: Choose the hard types}
We choose $\delta$ on the scale $T^{-1/2}$ so that a regret lower
bound of order $1/\delta$ becomes a $\sqrt T$ lower bound.
Its coefficient must also be large enough to account for the
correction $D_F$ in $u=D_F+\delta r$.
Specifically, set
$$
\begin{aligned}
H&:=32(D_F+1)\exp\{2L_\varepsilon(D_F+1)\},\\
\delta&:=\sqrt{H/T}.
\end{aligned}
$$
The constant $H$ is independent of $A$ and $T$.
For $T\geq\lceil H/d_c^2\rceil$, we have $\delta\leq d_c$,
so both $\theta_-$ and $\theta_+$ belong to $\Theta_\varepsilon$.

Suppose, toward a contradiction, that $r\leq1/\delta$.
Then $0\leq u\leq D_F+1$.
Since $u\mapsto16u e^{2L_\varepsilon u}$ is increasing for $u\geq0$,
\eqref{eq:catr-lower-testing} implies
$$
\begin{aligned}
H=T\delta^2
&\leq16u e^{2L_\varepsilon u}\\
&\leq16(D_F+1)e^{2L_\varepsilon(D_F+1)}
=\frac H2,
\end{aligned}
$$
which is impossible.
Thus every admissible policy incurs regret greater than
$1/\delta=\sqrt{T/H}$ for at least one of the two feasible types.

The pair of types is chosen independently of the policy.
Taking the infimum over $A\in\mathcal A$ therefore gives
$$
\begin{aligned}
\mathcal R_T^*(\Theta_\varepsilon)
&\geq\inf_{A\in\mathcal A}
\max_{\theta\in\{\theta_-,\theta_+\}}
\operatorname{Reg}_T(A;\theta)\\
&\geq\frac{\sqrt T}{\sqrt H}.
\end{aligned}
$$
This proves the $\Omega(\sqrt T)$ lower bound.
Together with CAT-R's upper bound, it establishes
$\mathcal R_T^*(\Theta_\varepsilon)=\Theta(\sqrt T)$
for every fixed prior and interior serving cost under the stated
assumptions.
\end{proof}